\documentclass[11pt,a4paper,onecolumn,preprintnumbers,amsmath,amssymb,superscriptaddress,aps]{revtex4}
\pdfoutput=1
\usepackage{graphicx}
\usepackage{dcolumn}
\usepackage{bm}
\usepackage{longtable}
\usepackage{cancel}
\usepackage{epsfig}
\usepackage{amssymb}
\usepackage{amsmath}
\usepackage{amsthm}
\usepackage{latexsym}
\usepackage{graphicx}
\usepackage{epstopdf}
\usepackage[normalem]{ulem} 
\usepackage{soul}
\usepackage{ulem}
\usepackage{tikz}
\usepackage{dsfont}
\usepackage[hyperfigures,breaklinks,colorlinks,citecolor=blue,linkcolor=red]{hyperref} 

\def\be{\begin{equation}}
\def\ee{\end{equation}}
\def\bearr{\begin{eqnarray}}
\def\eearr{\end{eqnarray}}
\def\bc{\begin{center}}
\def\ec{\end{center}}

\def\<{\bigl\langle}
\def\>{\bigr\rangle}

\def\Q{\mathrm{Q}}

\def\s{\sigma}

\def\nm{\nonumber}
\def\ul{\underline}

\begin{document}

\title{On one-step replica symmetry breaking in the Edwards-Anderson spin glass model}
\author{Gino Del Ferraro}
\affiliation{AlbaNova University Centre, KTH-Royal Institute of Technology Dept of Computational Biology, SE-106 91 Stockholm, Sweden }

\author{Chuang Wang}
\affiliation{State Key Laboratory of Theoretical Physics, Institute of Theoretical Physics, Chinese Academy of Sciences, Beijing 100190, China}
\affiliation{School of Engineering and Applied Sciences, Harvard University, 33 Oxford Street, Cambridge MA 02138 USA}

\author{Hai-Jun Zhou}
\affiliation{State Key Laboratory of Theoretical Physics, Institute of Theoretical Physics, Chinese Academy of Sciences, Beijing 100190, China}

\author{Erik Aurell}
\affiliation{AlbaNova University Centre, KTH-Royal Institute of Technology Dept of Computational Biology, SE-106 91 Stockholm, Sweden }
\affiliation{ACCESS Linnaeus Centre and Center for Quantum Materials, KTH-Royal Institute of Technology, SE-100 44 Stockholm, Sweden } 
\affiliation{Depts of Applied Physics and Computer Science, Aalto University, FIN-00076 Aalto, Finland}

\date{\today}

\begin{abstract}
We consider a one-step replica symmetry breaking description of the Edwards-Anderson spin glass model in 2D.
The ingredients of this description are a Kikuchi approximation to the free energy and a second-level
statistical model built on the extremal points of the Kikuchi approximation, which are also fixed
points of a Generalized Belief Propagation (GBP) scheme. 
We show that a generalized free energy can be constructed where
these extremal points are exponentially weighted by their Kikuchi free energy and a Parisi parameter $y$,
and that the Kikuchi approximation of this generalized free energy leads to second-level, one-step replica symmetry breaking (1RSB), GBP equations. 
We then proceed analogously to Bethe approximation case for tree-like
graphs, where it has been shown that 1RSB Belief Propagation equations admit a Survey Propagation solution. We discuss when and how the one-step-replica symmetry breaking GBP equations that we obtain also allow a simpler class of solutions which can be interpreted as a class of Generalized Survey Propagation equations for the single instance graph case. 
\end{abstract}

\maketitle

\section{Introduction}
\label{s:intro}
Low-temperature phases of systems with random interactions can be thought of as optimizing under a large set of highly interdependent and conflicting constraints. If a local optimum has been found it will typically be difficult to improve upon without adjusting many variables over a large domain, which is inherently difficult, for any optimization procedure. A physical process obeying detailed balance will furthermore care not directly about how different are two configurations but rather about how high are the energy barriers between them; for low enough temperature transition times should follow an Arrhenius law, $\tau_{transit}\sim \tau_{micro}e^{-\beta E_B }$, where $\tau_{micro}$ is a (microscopic) attempt time, $E_B$ is a barrier height and $\beta=\frac{1}{k_BT}$ is the inverse temperature. If barrier heights increase with system size $N$ different such low-lying states will be essentially disconnected for very long times (exponential in $N$), and the equilibrium Gibbs measure separates into a (weighted) sum over such states. 

Let the above-mentioned (metastable) states be labeled by $k$  and
each characterized, over some long time, by an average energy $E_{k}$ and a probability distribution $P_{k}$ with 
entropy $S_{k}$. They can then be considered as having (metastable) free energy densities 
$f_{k}=\frac{1}{N}\left(E_{k}-T S_{k}\right)$ 
and from these one can
construct a second-order statistical mechanics built on a second-order partition function 
\be \label{eq:GPF-2order}
\Xi(y) = \sum_{k} \textit{e}^{\,\, -\beta \, y N f_{k}} 
\ee
which is known as a one-step replica symmetry breaking (1RSB) scheme \cite{mezard2009information,zamponi2010mean,monasson1995structural,MezardParisiVirasoro}.
The Parisi parameter $y$ in (\ref{eq:GPF-2order}) plays the role of an inverse temperature
and is conjugate to a free energy-like quantity (referred to as a
generalized free energy) defined as $G(y) = - \frac{1}{\beta y} \ln \Xi$.
In the large-$N$ limit the sum in (\ref{eq:GPF-2order}) can be expected to be dominated by different subsets of states
for different $y$, and knowing the corresponding maximizing distributions gives, in analogy with 
standard statistical mechanics, information on the system in the
thermodynamic limit.

This program has been carried out with great success for dilute 
systems where the constraints are locally organized in tree-like
structures. 
For such networks the states can be found as the fixed points of the iterative scheme known as Belief Propagation
(BP)~\cite{YedidiaFreemanWeiss2003,WainwrightJordan2008}, and the free
energy of each state that enter in (\ref{eq:GPF-2order}) is the
corresponding Bethe approximation to the free energy.
Furthermore, the computation of (\ref{eq:GPF-2order}) can be carried out
by an iterative scheme called Survey Propagation (SP), 
formally quite similar to BP
\cite{MezardParisi2001,MezardParisiZecchina2002,mezard2002random,braunstein2005survey,mezard2009information}.

While there are several routes to SP theory,
we will here only be concerned with a generalization of one 
presented in \cite{mezard2009information, zamponi2010mean}
where the three steps are
\textit{(i)} an efficient computation of $f_{k}$ using the Bethe approximation, 
\textit{(ii)} an approximation of (\ref{eq:GPF-2order}) by a Bethe approximation on the second order, and 
\textit{(iii)} a reduction of this approximation to SP, which can also be efficiently computed. 
In spin glass language an analysis based on BP is referred
to as the Replica Symmetric (RS) level, while an analysis based on SP
is referred to a One-Step Replica Symmetry breaking (1RSB) level.
We will use the term 1RSB to generally underline effects that follow
from a sum over states as in (\ref{eq:GPF-2order}) as opposed to
considering only one state.
As our focus here will be on systems in finite dimension and an
analysis built on the Kikuchi approximation, we will also use the
terms ``Kikuchi approximation on second level'' and ``second-level
GBP''
when discussing steps conceptually corresponding to \textit{(ii)}
above,  and
``Generalized Survey Propagation'' (GSP) for \textit{(iii)}.
We note that for the Kikuchi approximation and the analogy of Belief
propagation (step \textit{(i)} above)
the term Generalized Belief Propagation
(GBP) is well established \cite{YedidiaFreemanWeiss2003,yedidia2000generalized}.

The rationale for considering a generalization of SP of this type
which we call GSP is the simple
fact that systems in finite dimension are not arranged in tree-like structures. The above sketched approach is therefore, for a physical system, only valid as a mean-field theory and is typically also presented as such. There exists a considerable literature as to whether these approaches apply to finite-dimensional systems \textit{cf.}~\cite{FisherHuse1988,Hartmann1997,Middleton1999,Parisi2008} -- which we will not enter except to point out that the question of barriers, and their heights, is one of geometry and hence strongly dependent on dimensionality. Indeed, even for infinite-dimensional systems much less is known about the distribution of barrier heights than about the distributions of free energy minima, and (\ref{eq:GPF-2order}) may therefore, 
as pertaining to dynamically separate states, even in idealized cases be somewhat of an ansatz. 

The Kikuchi (or cluster) expansion was introduced to improve upon the
mean-field estimates of thermodynamic
quantities~\cite{Kikuchi1951,an1988note,pelizzola2005cluster}. Algorithmically
it can be turned into Generalized Belief Propagation
(GBP)~\cite{yedidia2001bethe, yedidia2000generalized,
  yedidia2005constructing} which, though considerably more complicated
than BP, can be used to estimate marginal probabilities on graphs with
short loops and has applications to inference in lattice systems
appearing in image processing
problems~\cite{WainwrightJordan2008}. The question then arises if the
local minima of Kikuchi free energy, computed as fixed points of GBP,
can be turned into a second-level statistical mechanics as in (\ref{eq:GPF-2order}), and if that can be efficiently computed in analogy to SP. If that would be the case we could arrive at better systematic approximations to thermodynamics of finite-dimensional random systems and, perhaps conceptually more important, give a different kind of argument in favour of the spin glass approach in finite-dimensional system. This general issue has been already investigated in the recent literature  for both the single instance and averaged case ~\cite{rizzo2010replica,Lage-Castellanos2013} by using a different approach based on replicas which leads to a different class of equations. The differences between our and this approach will be discussed in Section \ref{sec:compare_rcv}.

In this paper we attempt to address the problem by directly writing a second-order partition function analogous to (\ref{eq:GPF-2order}) for the Edwards-Anderson spin glass model in two dimensions, but using instead of Bethe approximation a Kikuchi expansion based on regions containing one, two or four spins (``vertices'', ``rods'' and ``plaquettes''). This leads to a quite complex statistical model on second level, where the variables are GBP terms 
(``GBP messages'') obeying hard-core constraints (``GBP fixed point equations'') and
weighted by terms ${e}^{\,\, -\beta \, y N f_{k}}$ where the $f_{k}$ are the Kikuchi free energies computed from the fixed points of GBP.
We show how to perform a Kikuchi expansion for this second-level model using regions isomorphic to the regions on the first level model,
and hence also interpretable as (second-level) ``vertices'', ``rods'' and ``plaquettes''. In contrast to first level the amount of variables
in each second-level region is however large, as will be described in some detail below.
We also point out, as far as we know for the first time, that a Kikuchi expansion can be performed in many ways for such a second-level model,
our choice is only one of the simplest.
We are thus able to carry through the second step in the approach described above arriving at a definite -- though complicated -- set of equations which can be 
interpreted as 1RSB GBP equations which are not the same class of those obtained in \cite{rizzo2010replica} by the replica Cluster Variation Method (CVM), see discussion in Sec \ref{sec:compare_rcv}.

We then follow the same path which leads to SP, \emph{i.e} assume the
same ansatz, and discuss how the second-level GBP equations
which we obtain here also admit a simpler class of solutions. These
new equations can be interpreted as a generalized form of the Survey
Propagation equations, 
in short a form of Generalized Survey propagation (GSP). We believe that this derivation is of interest for treating 
finite-dimensional systems in spin glasses as well as for computer science and satisfiability related problems. 
Although theoretically consistent, these equations however appear hard
to solve computationally due to their large dimensionality and a product of different distributions as already pointed out in~\cite{rizzo2010replica}.
The purpose of the current work is thus to show an alternative approach to a generalized SP theory based on a new statistical model, in the same spirit as done in \cite{mezard2009information,zamponi2010mean} for the Survey Propagation approach. 

The paper is structured as follows. In
Section~\ref{s:Edwards-Anderson} we recall the Edwards-Anderson model and
in Section~\ref{s:GBP} we describe a Kikuchi approximation for this model based 
on regions ``vertices'', ``rods'' and ``plaquettes'', leading to a Generalized
Belief Propagation scheme.  
This presentation follows earlier contributions by two of 
us~\cite{zhou2011partition,zhou2012region,wang2013simplifying} to
which we refer for some of the details.
In Section~\ref{s:GBP2} we introduce and discuss a second-level GBP scheme and
we present the 1RSB GBP equations at the
plaquette region-graph level of the Kikuchi approximation.  
  The differences between our approach and the
  the replica cluster variational approach of \cite{rizzo2010replica} are
  discussed in this section.
In Section~\ref{s:no-GSP}, proceeding in analogy to the path which leads to 
the Survey Propagation algorithm, we show that this scheme does simplify
similarly to the SP case to a class of generalized Survey Propagation
equations. 
In Section~\ref{s:discussion} we sum up and discuss our results. Four appendices integrate the main text and in the last of them we discuss, to our knowledge for the first time, meaning and computation of determinant factors which appear both in the SP \cite{mezard2009information,zamponi2010mean} and generalized SP theory, constructed by using a second-level region graph.  

\section{The Edwards-Anderson model}
\label{s:Edwards-Anderson}
The 2D Edwards-Anderson (EA) model \cite{edwards1975theory} is defined on a finite dimensional square lattice by the Hamiltonian $H= -\sum_{(ij)}J_{ij}\s_i \s_j - \sum_i h_i \s_i$, where the first sum is over neighbouring dihcotomic spin variables $\s_i=\pm1$, $h_i$ is a local external field and the $J_{ij}$ are quenched random variables. Using a factor graph representation with $N$ variable nodes ($i=1,2,\dots,N$) representing the spins and $M$ factor nodes ($a=1,2,\dots, M$) representing the interactions between the spins, the Hamiltonian can alternatively be written as:
\be\label{eq:hamiltonian}
H(\ul \s)=  \sum_a^M E_a(\ul\s_{\partial a }) + \sum_i^N E_i(\s_i)
\ee
where $\ul \s \equiv (\sigma_1, \sigma_2, \ldots,
\sigma_N)$ denotes a generic spin configuration, and by
$\ul\s_{\partial a}$ we mean $\{\s_i | i \in \partial a \}$, the set
of variables which are involved in interaction $a$. 
In the case of a two dimensional lattice this is simply $\ul\s_{\partial a} = \{\s_i, \s_j\}$, but the more general notation will be convenient later. In the same representation the partition function can be written as:
\be \label{eq:Z}
Z(\beta)\equiv \sum_{\underline{\s}} {\textit{e}}^{-\beta H(\ul{\s})} = \sum_{\ul\s} \prod_{i=1}^N \psi_i (\s_i) \prod_{a=1}^M \psi_a(\ul \s_{\partial a})
\ee
where $\beta= 1/ k_B T$ is an inverse temperature and the terms
\be
\label{eq:potential}
\psi_i (\s_i) \equiv e^{\beta h_i \s_i}, \qquad  \psi_a(\ul\s_{\partial a}) \equiv e^{\beta J_{ij}^{(a)}\s_i \s_j}
\ee
are called ``potential functions'' in BP terminology. 
That interaction is in the more general scheme denoted $(a)$ hence for
later convenience we include that label also for the interaction coefficient also $J_{ij}^{(a)}$.
The equilibrium free energy is related to the partition function as
\begin{equation}
F(\beta)  \equiv -\frac{1}{\beta} \ln Z(\beta).
\label{eq:freeenergy2}
\end{equation}

       The EA model has been extensively studied in the 
         literature with the spin coupling constants $J_{i j}$ following the
         $\pm J$ bimodal distribution or the continuous Gaussian distribution
         with zero mean and variance $J^2$. If all the external magnetic fields
         $h_i=0$, this model is in the spin glass phase when the inverse 
         temperature $\beta$ exceeds certain critical value $\beta_c$. The value
         of $\beta_c$ is finite for systems in $D\geq 3$ dimensions, while
         $\beta_c = \infty$ for $D=2$. The spin glass transition
         therefore occurs
         at zero temperature for two-dimensional systems
         \cite{morgenstern1980magnetic,saul1993exact,jorg2006strong,thomas2011zero,thomas2009exact}.       

\section{Region graph representation and generalized belief propagation equations}
\label{s:GBP}
In this section we introduce a region graph description of the 2-dimensional lattice and use the Cluster Variation Method (CVM) as presented by Kikuchi \cite{Kikuchi1951,an1988note,pelizzola2005cluster} to compute 
an approximation to the free energy  and derive generalized belief propagation equations as presented in \cite{yedidia2005constructing,zhou2012region,wang2013simplifying}. This method was invented to make corrections to the Bethe free energy and therefore, algorithmically, to the BP method for loopy graphs. Alternative approaches have been carried out using loop corrections, both for binary and non-binary variables, in  \cite{chertkov2006loop,chernyak2007loop,montanari2005compute} and more recently in \cite{mori2015loop}.\\
A region-graph description of a lattice model defines a set $R$ of regions that include some chosen basic clusters of nodes, their intersections, the intersection of their intersections, and so on. The basic idea underlying the CVM is to approximate the free energy 
as a sum of the free energies of these basic clusters of nodes, minus the free energy of over-counted cluster intersections, minus the free energy of the over-counted intersections of intersections, and so on \cite{yedidia2000generalized}. Following the description in \cite{yedidia2005constructing}, we define a region average energy as $U_\alpha(P_\alpha) = \sum_{\ul{\s}_{\alpha}} P_{\alpha}(\ul{\s}_{\alpha}) E_\alpha(\ul{\s}_{\alpha})$, where $E_\alpha(\underline{\s_\alpha}) $ is the energy term of the region $\alpha$ similar to those appearing in equations \eqref{eq:hamiltonian}, and a region entropy $S_\alpha(P_\alpha)=- k_B\sum_{\ul{\s}_\alpha}P_\alpha(\ul{\s}_\alpha)\log P_\alpha(\ul{\s}_\alpha)$, in which ${\ul \s}_\alpha \equiv \{\sigma_i | i \in \alpha\}$ denotes a microscopic configuration of the same region.  The region probability $P_\alpha$ is a variational parameter and has to be determined from minimization condition. Let us observe that, in cases where $P_\alpha$ is the exact probability distribution of the region $\alpha$, the average energy correspond to the exact value whereas, differently, the region entropy remains an approximation \cite{yedidia2000generalized}. The Kikuchi free energy functional is then introduced as the sum of all the region contributions:
\begin{equation}
  \label{eq:F-Kikuchi}
  F_K = \sum_{\alpha}c_{\alpha} \sum_{\ul{\s}_{\alpha}} P_{\alpha}(\ul{\s}_{\alpha})   
  \biggl[E_\alpha(\ul{\s}_{\alpha}) + \beta^{-1}\log P_{\alpha}(\ul{\s}_{\alpha}) \biggr] 
\end{equation}
Above $E_\alpha(\ul{\s}_{\alpha})$ is the energy term of the region $\alpha$, in the EA model, for instance, $E_\alpha(\ul{\s}_{\alpha})=\sum_{i\in \alpha} h_i\s_i+ \sum_{a\in \alpha} J^{(a)}_{ij}\s_i\s_j$.
In \eqref{eq:F-Kikuchi} the region graph coefficients $c_{\alpha}$ are counting number assigned to each region to ensure that each term only contributes once \cite{wang2013simplifying} and constructed recursively by:
\be
\label{eq:c_coeff}
c_{\gamma}= 1- \sum_{\{\alpha:\, \alpha > \gamma\}} c_{\alpha}\,,
\ee
where with $\alpha > \gamma$ we indicate that $\alpha$ is an ancestor of $\gamma$, \textit{i.e.} the regions $\gamma$ is contained in the region $\alpha$. The index $\alpha$ in \eqref{eq:F-Kikuchi} can in principle run over all the regions in the region graph although keeping all of them would be cumbersome. 
A Kikuchi approximation of the free energy at the $R$-level then fixes a set $R$ of clusters made of maximal clusters and all their sub-clusters, and truncates the expansion of the free energy \eqref{eq:F-Kikuchi} retaining only terms corresponding to clusters in $R$. The CVM aims to the minimization of the above free energy \eqref{eq:F-Kikuchi} under the constraints that the marginals $P_{\alpha}$ are consistent. The saddle point equations derived from this minimization are the so called Generalized Belief Propagation equations. A detailed derivation of them, following this standard approach, is provided in Appendix~\ref{a:Kikuchi-derivation}.

\begin{figure}[!t]
  \begin{center}
    \includegraphics[width=6cm]{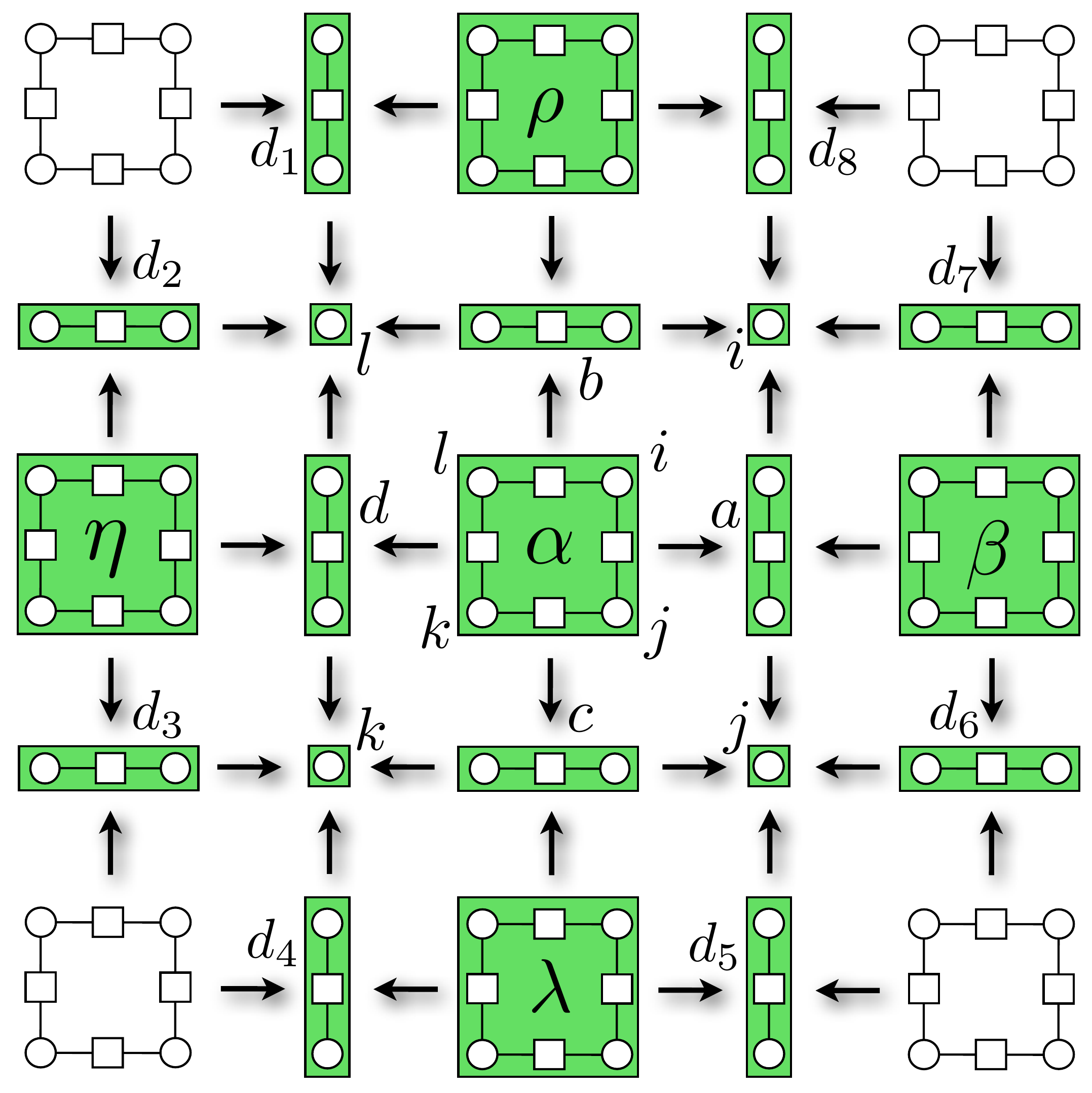}\hspace{1.5cm}
    \includegraphics[width=6cm]{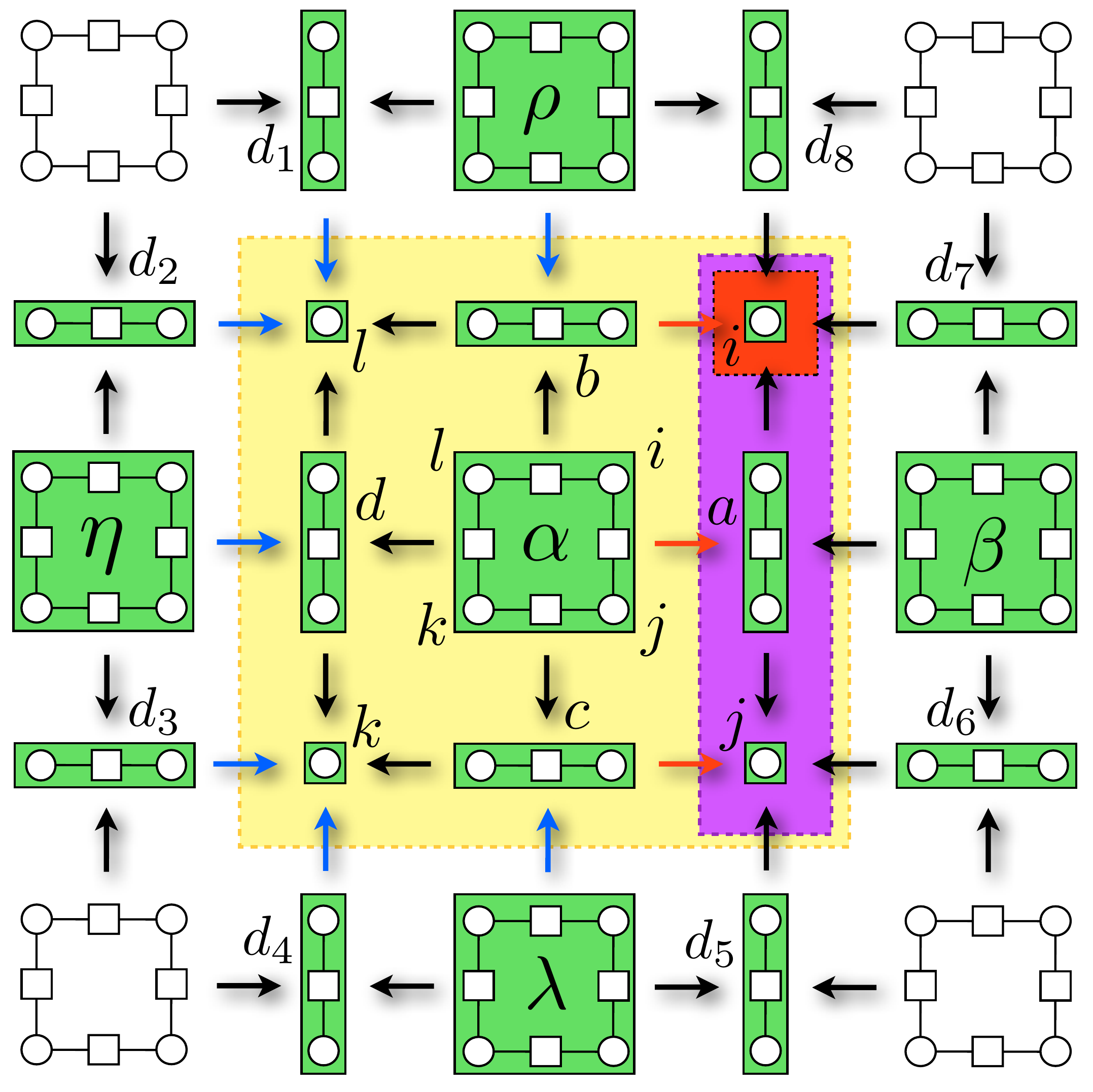}
    \caption{\footnotesize Left panel: Region graph representation of a 2D EA
      model in terms of plaquettes, rods and vertices. 
      Black arrows represent parent-to-child relations
        and are associated with messages from parents to child
      ($m_{\mu \to \nu}$ in the main text). Right panel: The squared yellow region represents the union of the border ($B_{\alpha}$) and interior ($I_{\alpha}$) of the $\alpha$ plaquette. The messages inside it belong to $I_{\alpha}$ whereas the messages that cross the border belong to $B_{\alpha}$. The purple rod region represents the same object for the rod $a$. Blue and red arrows represent the messages involved in the GBP equation from plaquette $\alpha$ to rod $a$, namely eq. \eqref{eq:GBPplaq} in the main text. The counting numbers for both figures are $c_\alpha=1$ for plaquettes, $c_a= -1$ for rods and $c_i=1$ for vertexes.}
    \label{fig:1}
  \end{center}
\end{figure}

An alternative formulation of the Kikuchi approximation and a derivation of GBP equations can be obtained formulating a region-graph description in terms of factor graphs and messages, \textit{i.e.} normalized probability distributions, between neighbouring regions. In what comes next we follow such a formulation as described in \cite{wang2013simplifying} in order to obtain consistency relations among regions which will be useful in the  derivation of the 1RSB Generalized Belief Propagation equations (see Sec. \ref{s:GBP2}). Within this description, generally a region $\mu$ can be a parent of another region $\nu$ if the set $\nu$ is contained in the set $\mu$, and when this is the case we indicate the relation as  $\mu \to \nu$. Hereafter we consider the squares or ``plaquettes'' to be the larger cluster $R$ and therefore we deal with three types of clusters or regions: plaquettes, rods and vertices (see Fig. \ref{fig:1}), although the description presented next is very general and not restricted to such maximal clusters. Each square contains four vertices and four interactions and is a parent of four rod regions; each rod region contains two vertices and one interaction and is a parent of two vertices; while each vertex region contains only one vertex and has no children. In Fig.~\ref{fig:1}, left panel, the parent-to-child relations $\mu \to \nu$ are represented by black arrows for the regions highlighted in green, namely plaquette, rods and vertices.
We recursively define an ancestor of a region to be a parent or the parent of an ancestor: a vertex thus has four parents (four rods) and four other ancestors (four plaquettes). Similarly we recursively define a descendant of a region to be a child or the child of a descendant. A plaquette thus has four children (four rods) and four other descendants (four vertices). A region graph $R$, within this description, is then a collection of all the regions
and the specified parent-to-child relations (\textit{i.e.}, arrows) between these regions.

Following a factor region graph description as in \cite{zhou2012region,wang2013simplifying}, the partition function \eqref{eq:Z} can then be written as 
\begin{equation}
\label{eq:Z_part}
    Z(\beta) = \sum\limits_{\underline{\s}}
    \prod_{\alpha \in R} \biggl[
    \prod_{i\in \alpha} \psi_i(\s_i) \prod_{a\in \alpha}
\psi_a(\underline{\s}_{\partial a})
    \biggr]^{c_\alpha} \; .
\end{equation}
where $\alpha$ in principle ranges over all the possible regions. As noted in \cite{wang2013simplifying} it is possible to introduce sets of messages $m_{\mu \to \nu}(\ul\s_{\nu})$, \textit{i.e.} normalized probability distributions, between parent-child regions in $R$, with the only constraint that $\sum_{\s_{\nu}} m_{\mu \to \nu}(\ul\s_{\nu})=1$. Indeed we observe that for each direct edge $\mu \to \nu$ the following relation holds 
\begin{equation}
\label{eq:c-identity}
\sum\limits_{\{\alpha \ : \ \mu \in B_\alpha, \, \nu \in I_\alpha\}} c_\alpha
= \sum\limits_{\alpha \geq \nu} c_\alpha -
\sum\limits_{\alpha \geq \mu} c_\alpha = 1-1 = 0\, ,  
\end{equation}
where we have for each region $\alpha$ defined $I_\alpha \equiv \{ \gamma: \gamma \leq \alpha\}$
to be the set formed by the region $\alpha$ and all its descendants, and $B_\alpha$, the ``boundary'' of $\alpha$,
\textit{i.e.} to be the set of regions not belonging to $I_\alpha$ but parental to at least one region in $I_\alpha$ \cite{wang2013simplifying} (see Fig \ref{fig:1}, right panel).
Therefore the partition function \eqref{eq:Z_part} can be rewritten as:
\begin{equation}
Z(\beta) = \sum\limits_{\underline{\s}}
\prod\limits_{\alpha\in R}
\biggl[ 
\prod\limits_{i\in \alpha} \psi_i(\s_i)
\prod\limits_{a\in \alpha} \psi_a(\underline{\s}_{\partial a})
\prod\limits_{\{\mu\rightarrow \nu \
: \
\mu\in B_\alpha, \nu\in I_\alpha
\}} m_{\mu\rightarrow \nu}(\underline{\s}_\nu)
\biggr]^{c_\alpha}  \; 
\label{eq:Z-exp-2}
\end{equation}
indeed due to relation \eqref{eq:c-identity} each message appears in \eqref{eq:Z-exp-2} is net total zero times. 
A Kikuchi approximation of this partition function corresponds  to take into account only
some chosen set of maximal clusters $R$ in  \eqref{eq:Z-exp-2}, their intersections, the intersection of their intersections, and so on. The Kikuchi approximation of the above partition function then reads as \cite{zhou2012region}
\begin{equation}
Z_K(\beta) =
\prod\limits_{\alpha\in R}
\biggl[ \sum\limits_{\underline{\s}_\alpha}
\prod\limits_{i\in \alpha} \psi_i(\s_i)
\prod\limits_{a\in \alpha} \psi_a(\underline{\s}_{\partial a})
\prod\limits_{\{\mu\rightarrow \nu \
: \
  \mu\in B_\alpha, \ \nu\in I_\alpha
  \}} m_{\mu\rightarrow \nu}(\underline{\s}_\nu)
\biggr]^{c_\alpha}  \; 
\label{eq:Z_Kikuchi}
\end{equation}
where, differently from \eqref{eq:Z-exp-2}, here $\alpha$ ranges over all regions up to a maximal cluster $R$, for instance restricted to plaquettes, rods and vertices for the case discussed here. The corresponding Kikuchi approximation of the free energy in terms of messages  $m_{\mu \to \nu}(\ul\s_{\nu})$, reads as:
\begin{equation}
F_K(\beta) = -\beta^{-1}
\sum\limits_{\alpha\in R} c_{\alpha} \log
\biggl[ \sum\limits_{\underline{\s}_\alpha}
\prod\limits_{i\in \alpha} \psi_i(\s_i)
\prod\limits_{a\in \alpha} \psi_a(\underline{\s}_{\partial a})
\prod\limits_{\{\mu\rightarrow \nu \
: \
  \mu\in B_\alpha, \ \nu\in I_\alpha
  \}} m_{\mu\rightarrow \nu}(\underline{\s}_\nu)
\biggr] \; .
\label{eq:F-Kikuchi-2}
\end{equation}
Analogously to the standard derivation of the CVM, we then require this free energy to be stationary respect to a chosen set of probability functions $\{m_{\mu \to \nu}\}$, namely:
\begin{equation}\label{eq:dF0}
\frac{\delta F_K}{\delta m_{\mu\rightarrow \nu}} = 0 \; ,\quad\quad\quad
\forall \,\,(\mu\rightarrow \nu) \in R \; .
\end{equation}
As noted in \cite{zhou2012region,wang2013simplifying}, this equation is satisfied if the consistency condition between parent-child regions of two marginal probability distributions is satisfied, namely
\begin{equation}
\label{eq:consist}
\sum\limits_{\underline{\s}_\mu \backslash \underline{\s}_\nu} \omega_\mu(
\underline{\s}_\mu) \ =
\ \omega_\nu(\underline{\s}_\nu) \; ,
\end{equation}
where $\omega_\mu$ is a Boltzmann factor 
\begin{equation}
\label{eq:omega}
\omega_\alpha(\underline{\s}_\alpha)
  \equiv \frac{1}{z_\alpha} \prod\limits_{i\in \alpha} \psi_i(\s_i)
\prod\limits_{a\in \alpha} \psi_a(\underline{\s}_{\partial a})
\prod\limits_{\{\mu\rightarrow \nu \ : \ \mu \in B_\alpha, \nu \in I_\alpha\}}
m_{\mu\rightarrow \nu}(\underline{\s}_\nu) \; 
\end{equation}
called \emph{belief} of the region $\alpha$ in the message-passing language and $z_\alpha$ is a normalization factor associated to the same region. Equation \eqref{eq:consist} ensures that the marginal probability distribution $\omega_\mu(\underline{\s}_\mu)$ and $\omega_\nu(\underline{\s}_\nu)$ of each parent-to-child pair come from the same joint probability distribution. The constraints on the marginals, together with \eqref{eq:omega}, lead to the generalized belief-propagation equation on each directed edge $\mu \to \nu$ of the region graph $R$
\begin{eqnarray}
& & \hspace*{-1.0cm} \prod\limits_{\{\substack{
\alpha \rightarrow \gamma \ : \\
\alpha \in B_\nu \cap I_\mu, \gamma \in I_\nu
\}}}
m_{\alpha \rightarrow \gamma}(\underline{\s}_\gamma )
=  C
\sum\limits_{\underline{\s}_\mu \backslash
\underline{\s}_\nu} \prod\limits_{j\in \mu \backslash \nu} \psi_j(\s_j)
\prod\limits_{b\in \mu \backslash \nu} \psi_b(\underline{\s}_{\partial b})
\prod\limits_{\substack{\{\eta \rightarrow \tau \ : \\ \eta \in B_\mu ,
\tau \in I_\mu \backslash I_\nu\}}}
 m_{\eta\rightarrow \tau}(\underline{\s}_\tau) \; 
 \label{eq:gbp}
\end{eqnarray}
where $C$ is a normalization constant determined by $\sum_{\ul\s_\nu}m_{\mu \to \nu}(\ul\s_\nu)=1$. 

\subsection{Plaquette-rod-vertex region graph description}

As a concrete example and as a reference for the following sections, we explicitly give the Kikuchi partition function (free energy) and the corresponding generalized belief propagation equations for a region graph description of the 2D Edward-Anderson model. Letting now Greek letter $\alpha$ stand for plaquettes, $a$ for rods and $i$ for vertices and introducing auxiliary functions
\begin{align}\label{eq:Zalpha}
Z_\alpha &= \sum\limits_{\underline{\s}_\alpha} \prod_{i\in \alpha} \psi_i(\s_i) \prod_{a\in \alpha} \psi_a(\underline{\s}_{\partial a}) \prod_{\{\mu\rightarrow \nu \: \ \mu\in B_\alpha,\, \nu\in I_\alpha \}} m_{\mu\rightarrow \nu}(\underline{\s}_\nu), \\ \label{eq:Za}
Z_a &= \sum\limits_{\underline{\s}_a} \psi_a(\underline{\s}_{\partial a}) \prod_{i\in a} \psi_i(\s_i) \prod_{\{\mu\rightarrow \nu \: \ \mu\in B_a, \, \nu\in I_a \}} m_{\mu\rightarrow \nu}(\underline{\s}_\nu),\\ \label{eq:Zi}
Z_i &= \sum\limits_{\s_i}  \psi_i(\s_i)  \prod_{\{\mu\to i: \ \mu\in B_i \}} m_{\mu\to i}({\s}_i),
\end{align}
the Kikuchi partition function \eqref{eq:Z_Kikuchi} can be written compactly as
\be
Z_K = e^{-\beta F_K} = \prod_{\alpha \in R} Z_{\alpha} \prod_{a \in \alpha} Z_a^{-1} \prod_{i \in a} Z_i
\label{eq:K}
\ee
where $F_K$ refers to the Kikuchi free energy \eqref{eq:F-Kikuchi-2}. 
As messages go from parents to children, and as in the chosen region graph we have only parents (plaquettes), children (rods) and grand-children (vertices), we have only two instances of the general equation (\ref{eq:gbp}). With reference to Fig. \ref{fig:1}, right panel, for the labeling, the first is plaquettes-to-rods and reads
\begin{align}\nm
m_{\alpha \to a}(\s_i, \s_j)m_{b\to i}(\s_i)m_{c\to j}(\s_j) &= \frac{1}{Z_{\alpha \to a}} \sum_{\s_l,\,\s_k} \psi_l(\s_l)\psi_k(\s_k)\psi_b(\s_l,\s_i)\psi_d(\s_l,\s_k)\psi_c(\s_k,\s_j)\\ \nm
&\times m_{\rho \to b}(\s_i, \s_l) m_{d_1 \to l}(\s_l) m_{d_2 \to l}(\s_l) m_{\eta \to d}(\s_l, \s_k) \\
&\times m_{d_3 \to k}(\s_k)m_{d_4 \to k}(\s_k) m_{\lambda \to c}(\s_k, \s_j)
\label{eq:GBPplaq}
\end{align}
where $Z_{\alpha \to a}$ is a normalization factor. 
The messages which appear on the LHS of
(\ref{eq:GBPplaq}) are those depicted with red arrows in Fig \ref{fig:1} (right panel) whereas those on the RHS are depicted as blue arrows. 
Equation \eqref{eq:GBPplaq} can be written compactly, in a form we will use later, as
\be\label{eq:GBPcompact1}
m_{\alpha \to a}(\ul\s_a) = \mathcal{F}_{\alpha \to a}\big(\{ m_{\eta\rightarrow \tau}\}_{[\eta \in B_\alpha,
\tau \in I_\alpha \backslash I_a]}, \{m_{\beta \to \gamma}\}_{[\beta \in (B_a \cap I_\alpha)\backslash \alpha, \gamma \in I_a \backslash a]} 	\big)
\ee
The second instance is the rods-to-vertices equation and reads
\begin{align}
m_{a \to i}(\s_i) = \frac{1}{Z_{a \to i}} \sum_{\s_j} \psi_{j}(\s_j) \psi_{a}(\s_i, \s_j) m_{\alpha \to a}(\s_i,\s_j)\, m_{\beta \to a}(\s_i,\s_j) m_{c \to j}(\s_j)m_{d_5 \to j}(\s_j)m_{d_6 \to j}(\s_j)
\label{eq:GBProds-main}
\end{align}
where $Z_{a \to i}$ is a normalization factor. Referring to Fig. \ref{fig:gbp_arrows} (right panel), the messages on the RHS of  \eqref{eq:GBProds-main} are those depicted in black whereas the message on the LHS is depicted in white.
 Similarly to \eqref{eq:GBPplaq}, equation \eqref{eq:GBProds-main} can also be written in a compact form as
\be\label{eq:GBPcompact2}
m_{a \to i}(\s_i) = \mathcal{F}_{a \to i}\big(\{ m_{\eta\rightarrow \tau}\}_{[\eta \in B_a,
\tau \in I_a \backslash I_i]} \big)
\ee
The GBP equations \eqref{eq:GBPplaq} and \eqref{eq:GBProds-main} have a somewhat abstract flavor. 
In practice they should be considered as iterative equations for the real parameters
describing the various probability distributions. For completeness we 
give such a more detailed description in Appendix~\ref{a:detailed-GBP} below. 
The region graph shown in Fig.~\ref{fig:1} is redundant in the sense that 
there are two directed paths from a plaquette region $\alpha$ to each of its
four vertex child regions \cite{zhou2012region}. Because of this redundancy the
LHS of the GBP equation (\ref{eq:GBPplaq}) is a product of three messages,
which leads to practical complications in updating the plaquette-to-rod message
$m_{\alpha \rightarrow a}(\sigma_i, \sigma_j)$. In general we regard a region
graph as a non-redundant region graph if there is at most one directed path 
from any region to any another region, otherwise the region graph is regarded as
redundant \cite{zhou2012region}. In the case of a non-redundant region graph
the GBP equation can be much simplified and the LHS of each GBP equation
contains only one message (see \cite{zhou2012region} for detailed
discussions). Although redundant region graphs bring in computational 
complications, the main reason to prefer redundant region graphs over 
non-redundant ones is that, in a redundant region graph, more consistency
constraints are enforced on the parent-to-child messages and consequently
the results are more accurate.

\subsection{Gauge invariance of the GBP equations}\label{Sec:GF}

\begin{figure}[!t]
\begin{center}
\includegraphics[width=4cm]{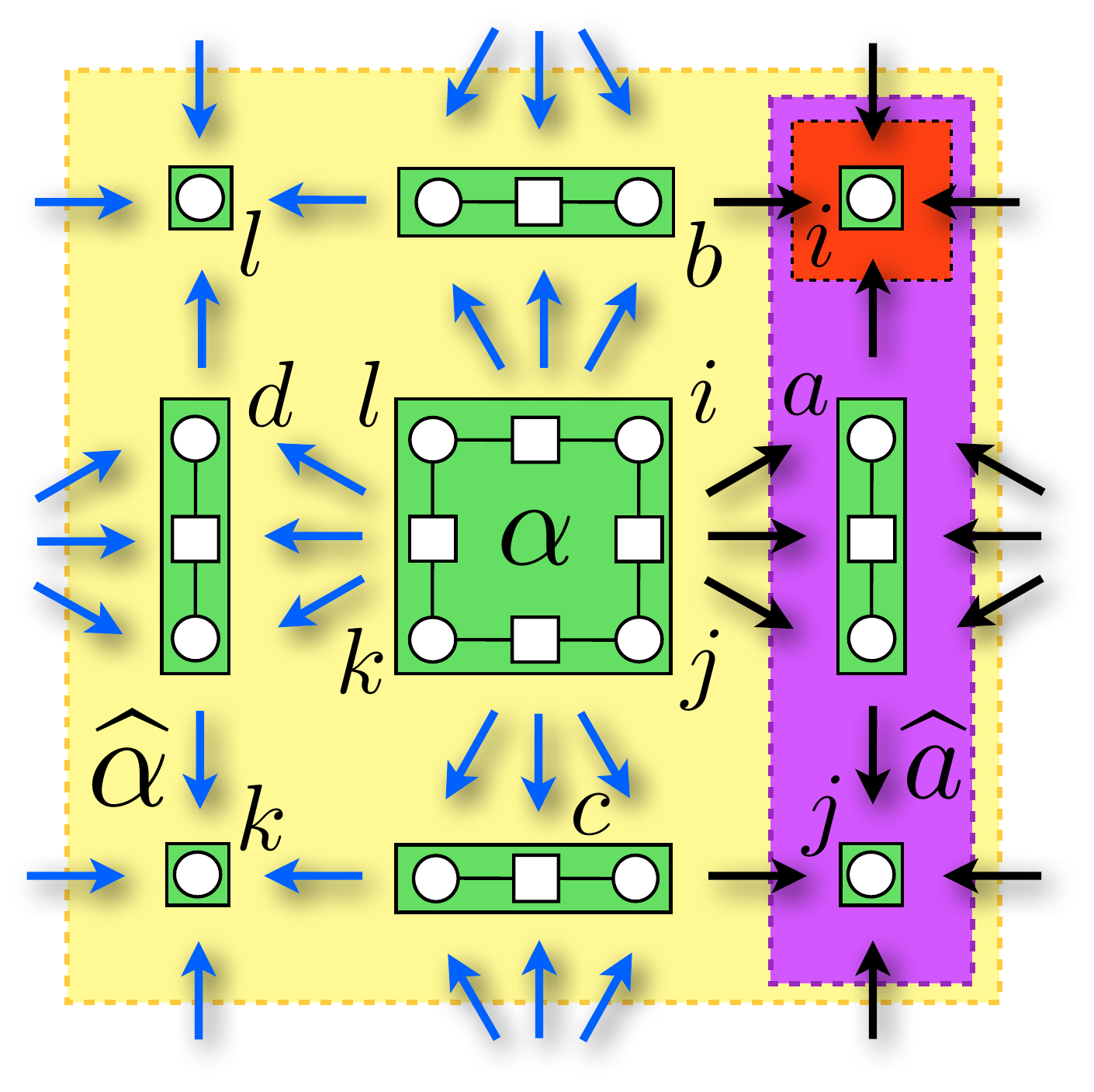}\hspace{1.5cm}
\includegraphics[width=1.6cm]{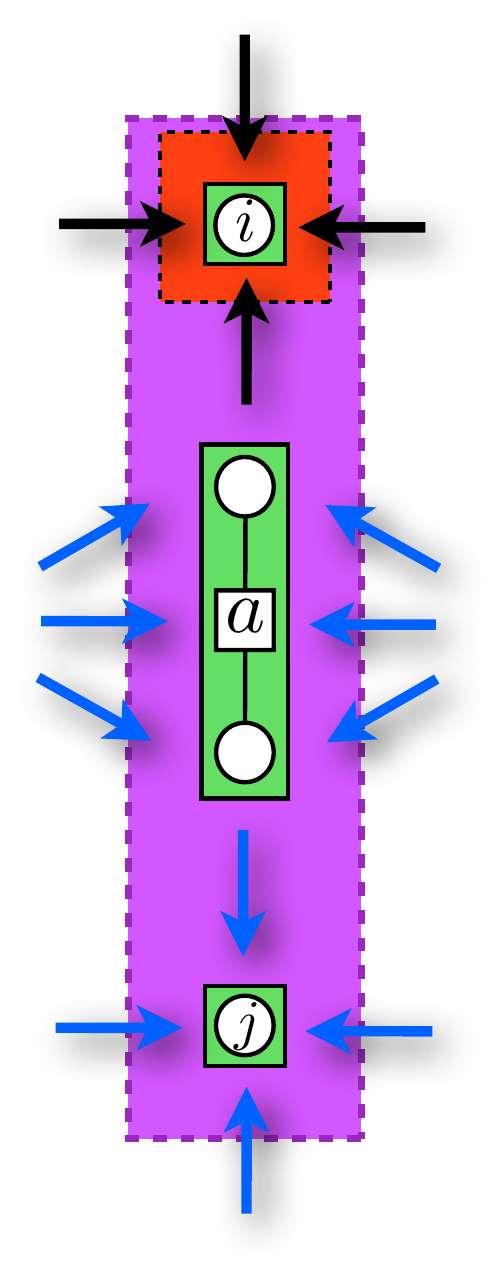}\hspace{1.5cm}
\includegraphics[width=1.6cm]{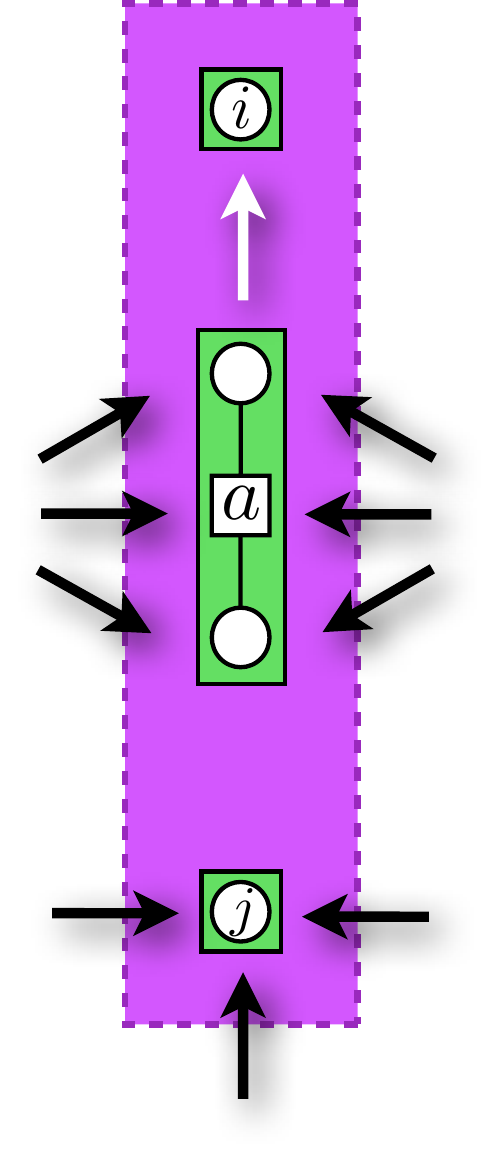}
\caption{\footnotesize Left panel: First and second level region graph description overlapped for the 2D EA model. Messages integrated over on the LHS of \eqref{eq:consistP} are illustrated with blue arrows whereas those on the RHS are pictured with black arrows. Central panel: same kind of illustration for equation \eqref{eq:consistR}. Right panel: pictorial representation of GBP rod-to-vertex equation \eqref{eq:GBProds-main}. The message on the LHS is pictured with a white arrow whereas black arrows illustrate the messages on the RHS.}
\label{fig:gbp_arrows}
\end{center}
\end{figure}

When the GBP equations are written to be iterated, for convenience, the messages are parametrized in terms of fields by using an exponential representation. In principle the messages can be left not-normalized therefore the following parametrization does not take into account their normalization which, when and if needed, can however be enforced at any time. Messages $m_{a \to i}(\s_i)$ are written as $e^{\beta u_{a \to i}\s_i}$ where  $u_{a \to i}$ is a real number known as the cavity bias on spin $i$ from interaction $a$. Similarly, messages $m_{\alpha \to a}(\s_i,\s_j)$ are written as $e^{\beta\left(u_{\alpha \to a}^{(i)}\s_i+u_{\alpha \to a}^{(j)}\s_j+U_{\alpha \to a}\s_i\s_j\right)}$ where the new quantity $U_{\alpha \to a}$ parametrizes cooperative interactions. The GBP equations \eqref{eq:GBPplaq} and \eqref{eq:GBProds-main} are then to be understood as relations between parameters of the types  $u_{a \to i}^{(i)}$, $u_{\alpha \to a}^{(i)}$ and $U_{\alpha \to a}$. In Appendix \ref{a:detailed-GBP} a derivation and the explicit expression for the GBP equations in terms of fields is reported. Here we just show their functional dependence which is the main thing for what follows. The link-to-vertex equation in term of fields reads as \cite{dominguez2011characterizing}
\be \label{eq:field_LV}
u_{a \to i} = \mathcal{L}_{a \to i}(u_{\alpha \to a}^{(i)} + u_{\beta \to a}^{(i)}, J_{ij}^{(a)} +U_{\alpha \to a } + U_{\beta \to a}, h_j +u_{\alpha \to a}^{(j)}+ u_{\beta \to a}^{(j)} + u_{c\to j} + u_{d_5 \to j} + u_{d_6 \to j} )
\ee
(see Appendix \ref{a:detailed-GBP} for the explicit formula). In Figure \ref{fig:gbp_arrows}, right panel, the message $u_{a \to i} $ on the LHS of \eqref{eq:field_LV} is illustrated as a white arrow whereas all the messages which appear on the RHS are illustrated as black arrows.  
The equations for the fields involved in the plaquette-to-link equations, using a similar notation as in \cite{dominguez2011characterizing}, are given by 
\begin{align} \label{eq:fields_PL1}
U_{\alpha \to a}=&\,\mathcal{F}_{\alpha \to a}(\vec U, \vec u) = \frac{1}{4 \beta} \log\Big[\frac{K(1,1)K(-1,-1)}{K(1,-1)K(-1,1)}\Big] \, ,\\ 
\label{eq:fields_PL2}
u_{\alpha \to a}^{(i)}= &\,\mathcal{G}_{\alpha \to a}^{(i)}(\vec U, \vec u)  =  u_{\rho \to b}^{(i)}- u_{b \to i} +\frac{1}{4 \beta} \log \Big[\frac{K(1,1)K(1,-1)}{K(-1,1)K(-1,-1)}\Big]\, ,\\
u_{\alpha \to a}^{(j)} =&\,\mathcal{G}_{\alpha \to a}^{(j)}(\vec U, \vec u) = u_{\lambda \to c}^{(j)}-  u_{c \to j}+ \frac{1}{4 \beta} \log\Big[\frac{K(1,1)K(-1,1)}{K(1,-1)K(-1,-1)}\Big] \, ,
\label{eq:fields_PL3}
\end{align}
where $K(\s_i,\s_j)$ is defined in the Appendix \ref{a:detailed-GBP} and only depends on the messages in $B(\alpha)$, which we here call ``external'' field-messages, \textit{i.e.} for the $\alpha \to a$ equations only those illustrated as blue arrows entering the yellow square region in Figure \ref{fig:gbp_arrows}, left panel. Therefore from \eqref{eq:fields_PL1} we observe that the field-messages of the kind $U_{\alpha \to a}$ do not depend on any messages in $I(\alpha)$ that we call ``internal'' message, whereas from \eqref{eq:fields_PL2} and \eqref{eq:fields_PL3} we note that the dependence on internal messages appears linearly only through the messages of the kind $u_{b \to i}$ and $u_{c \to j}$ respectively.

In \cite{dominguez2011characterizing} the authors point out that parent-to-child GBP algorithms, like the one we use here, posses a gauge invariance on the field-message values. In fact, GBP equations for such algorithms admit a freedom in the choice of the fields that has no effect on the fixed point solutions. Therefore, following \cite{dominguez2011characterizing}, every set of values of the fields which satisfies a fixed point can be changed by an arbitrary constant $\delta$ as
\begin{gather}
\begin{aligned}
u_{a \to i} \to u_{a \to i} + \delta, \qquad u_{\alpha \to a}^{(i)} \to u_{\alpha \to a}^{(i)} + \delta, \\
u_{b \to i} \to u_{b \to i} - \delta, \qquad u_{\alpha \to b}^{(i)} \to u_{\alpha \to b}^{(i)} - \delta, 
\end{aligned}\label{eq:Gauge}
\end{gather}
and the resulting set of values is still a solution of the GBP equations. This gauge invariance can be fixed in several ways, one of the simplest is to set to zero one of the four $u$-field in the equation \eqref{eq:Gauge}. This gauge fixing method will be used in the following sections when building a second-level statistical model based on the GBP fixed point solutions. 
%
%
%


\section{One-step replica symmetry breaking GBP equations}
\label{s:GBP2}
We here take up the approach discussed in the Introduction and
consider a second-order statistical model built on a second level
(1RSB) partition function \eqref{eq:GPF-2order}. In the picture we
want to consider here, the Gibbs measure is, in principle, not anymore a pure state but
rather decomposes into a convex linear combination of pure states 
\cite{monasson1995structural,mezard2009information,zamponi2010mean,MezardParisiVirasoro}.
We assume that in each such state labeled by $k$ the GBP equations \eqref{eq:gbp} are satisfied 
and we denote with $\{m_{\mu \to \nu}^{(k)}\}$ the set of their solution, for every $\mu \to \nu \in R$. This assumption is supported by the recent results in \cite{lage2014message} where it is shown that  metastable states found with Monte Carlo algorithm are linked to fixed points of the Cluster Variational Method. 
A Kikuchi approximation of the generalized free energy \eqref{eq:GPF-2order} is then given by replacing the exact free energy of every state by the Kikuchi free energy computed at the same state, \textit{i.e.} $N f_k \to F_K{\{m_{\mu \to \nu}^{(k)}\}}$ in \eqref{eq:GPF-2order}. The resulting  second-order partition function is defined as a sum over fixed points of GBP weighted by their respective Kikuchi free energy:
\be
\Xi(y) = \sum_{k} \textit{e}^{\,-\beta \, y N f_k} \simeq \sum_{k \in E} \textit{e}^{ -\beta\,y  F_K{\{m_{\mu \to \nu}^{(k)}\}}} \equiv \Xi_K(y) \; ,
\label{eq:second-order-GDF}
\ee
where $E$ denotes the set of GBP fixed points and with $\Xi_K$ we define the Kikuchi approximation of the replicated or second order partition function.
By standard folklore we expect either GBP to have only one fixed point, in which case the sum and
the additional parameter $y$ in (\ref{eq:second-order-GDF}) gives no new information,
or to have many fixed points, in which case their relative contributions to the sum are controlled by $y$.
The messages $m_{\mu \to \nu}$ are considered variables in a statistical model, and the restrictions
that the messages are fixed points of GBP are interpreted as hard-core constraints
\be\label{eq:Xi_K}
\Xi_K=  \sum_{k \in E} \, \int \mathcal{D}m_{\mu \to \nu}\,
\textit{e}^{\,-\beta\,  y  F_K\{m_{\mu \to \nu}\}}\, 
\prod_{\langle \mu \to\nu\rangle} \delta{[m_{\mu \to \nu}-m^{(k)}_{\mu \to \nu})]}
\ee
where on each pair of connected regions ${\langle \mu\to\nu\rangle}$,
plaquette to link and link to vertex, we have a message $m_{\mu \to
  \nu}$ and the delta functions enforce that these variables
correspond
to a given fixed point, labeled by $(k)$, of the GBP equations. 

Actually we cannot however sum over an enumeration of the fixed points
of GBP, that would not be any simplification of the problem.
Instead we must include the hard-core constraints inside the
integration as
\begin{align}
\Xi_K(y)&=\int \mathcal{D}m_{\mu \to \nu}\, \textit{e}^{\,-\beta\,  y
  F_K\{m_{\mu \to \nu}\}}\, 
\prod_{\langle    \mu\to\nu\rangle} 
\delta{[m_{\mu \to \nu}-\mathcal{F}_{\mu \to \nu}(\hat{\ul m})]}
\label{eq:noJacobian}
\end{align}
where $\hat{\ul m}$ indicates all the dependencies of the
function $\mathcal{F}$ from the messages involved in the GBP equations. 
By postulating (\ref{eq:noJacobian}) we have ignored Jacobians from integration of the delta functions.
If we insist that $\Xi(y)$ in \eqref{eq:second-order-GDF} is only a sum over fixed points, with no pre-factors, we should
more properly have
\begin{align}
\Xi_K(y)&=\int \mathcal{D}m_{\mu \to \nu}\, \textit{e}^{\,-\beta\,  y
  F_K\{m_{\mu \to \nu}\}}\,
\Big|\text{det}\Big[\frac{\partial(m_{\mu \to \nu} -\mathcal{F}_{\mu \to \nu})}{\partial m_{\mu' \to \nu'}}\Big]\Big|   
\prod_{\langle \mu\to\nu\rangle} \delta{[m_{\mu \to \nu}-\mathcal{F}_{\mu \to \nu}(\hat{\ul m})]}
\label{eq:Jacobian}
\end{align}
where the determinant is of a large matrix - coupling the first-level
variables (GBP messages) as they are connected by the hard-core
constraints (GBP update equations) - computed at the fixed point of the GBP equations.  
This value is independent on the Parisi parameter $y$ and therefore
contributes only to the entropic part of re-weighting term
${e}^{\,-\beta\,  y F_K\{m_{\mu \to \nu}\}}$ in equation
\eqref{eq:Jacobian}. For tractability we will from here neglect this
term, to discuss it again in Section \ref{s:discussion} and Appendix \ref{s:determinants}. We just mention that a similar determinant appears also for a tree-like topology and in that case it is similarly neglected as discussed in \cite{mezard2009information,zamponi2010mean}.
Let us also observe that the above integrals in
\eqref{eq:noJacobian} and \eqref{eq:Jacobian} converge only if the
gauge invariance of the GBP equation is fixed (see Section
\ref{Sec:GF}), otherwise a volume element of the
parametrization invariance appears. We assume here that this is the case and we specify in more details how to fix the gauge invariance at the second-level statistical model in Section \ref{sec:ConsEqPR}.
Using \eqref{eq:K} and \eqref{eq:noJacobian} we can then write
\be
\Xi_K= \sum_{k \in E} \textit{e}^{\,-\beta\,  y F_K{\{m_{\mu \to \nu}^{(k)}\}}} = \int \mathcal{D}m_{\mu \to \nu}\, \prod_{\alpha \in R}Z_\alpha^{\; y} \prod_{a \in R} Z_a^{-y} \prod_{i \in R} Z_i^{\;y}\Big[ \prod_{\langle \mu\to\nu\rangle} \delta{[m_{\mu \to \nu}-\mathcal{F}_{\mu \to \nu}(\hat{\ul m})]} \Big]
\label{eq:GPFfinal}
\ee
where we recall that $Z_\alpha$, $Z_a$ and $Z_i$ are functions of the
messages as given in \eqref{eq:Zalpha}, \eqref{eq:Za} and
\eqref{eq:Zi}. We will further show below that each $\delta$-function in
(\ref{eq:GPFfinal}) can be assigned to one or more regions on a
higher-level graph. 
We therefore interpret equation \eqref{eq:GPFfinal} as the partition function of
a new model where the messages represent the new variables, 
and the $Z$'s together with an appropriate set of $\delta$-functions
are the analogues on the second level of GBP potential functions on
the first level. We note that these second-level potential functions
contain both soft constraints in the  $Z$'s 
and hard constraints in the $\delta$-functions. 

\subsection{Higher-level region graph and second-level GBP}

In the previous section we introduced a second-level graphical model by introducing the new partition function \eqref{eq:GPFfinal}. The next step towards 1RSB GBP equations is to observe that the weighting terms $w_k = \exp{(-\beta y F_K\{m_{\mu\to\nu}^{(k)}\})}/ \Xi_K$ induce a distribution of the messages and that, furthermore, this distribution can be represented as a graphical model. In this new second-level region-graph model the new variables are messages and the factor nodes are new potential functions or constraints. Each factor node is connected to all the variable nodes on which it depends 
\textit{i.e.} for a potential functions to all 
the arguments of respectively $Z_\alpha$, $Z_a$ and $Z_i$, and for a constraint
$\delta{[U_{\mu \to \nu}-\mathcal{F}_{\mu \to \nu}(\hat{\ul U},\hat{\ul u})]}$ or $\delta{[u_{\mu \to \nu}-\mathcal{F}_{\mu \to \nu}(\hat{\ul U},\hat{\ul u})]}$
both to $U_{\mu \to \nu}$ or $u_{\mu \to \nu}$ and to all the arguments of $\mathcal{F}_{\mu \to \nu}$.
This auxiliary model is a graph expansion of the original model, as discussed for the case of BP in~\cite{mezard2009information, zamponi2010mean}.

\begin{figure}[!t]
\begin{center}
\includegraphics[width=6.2cm]{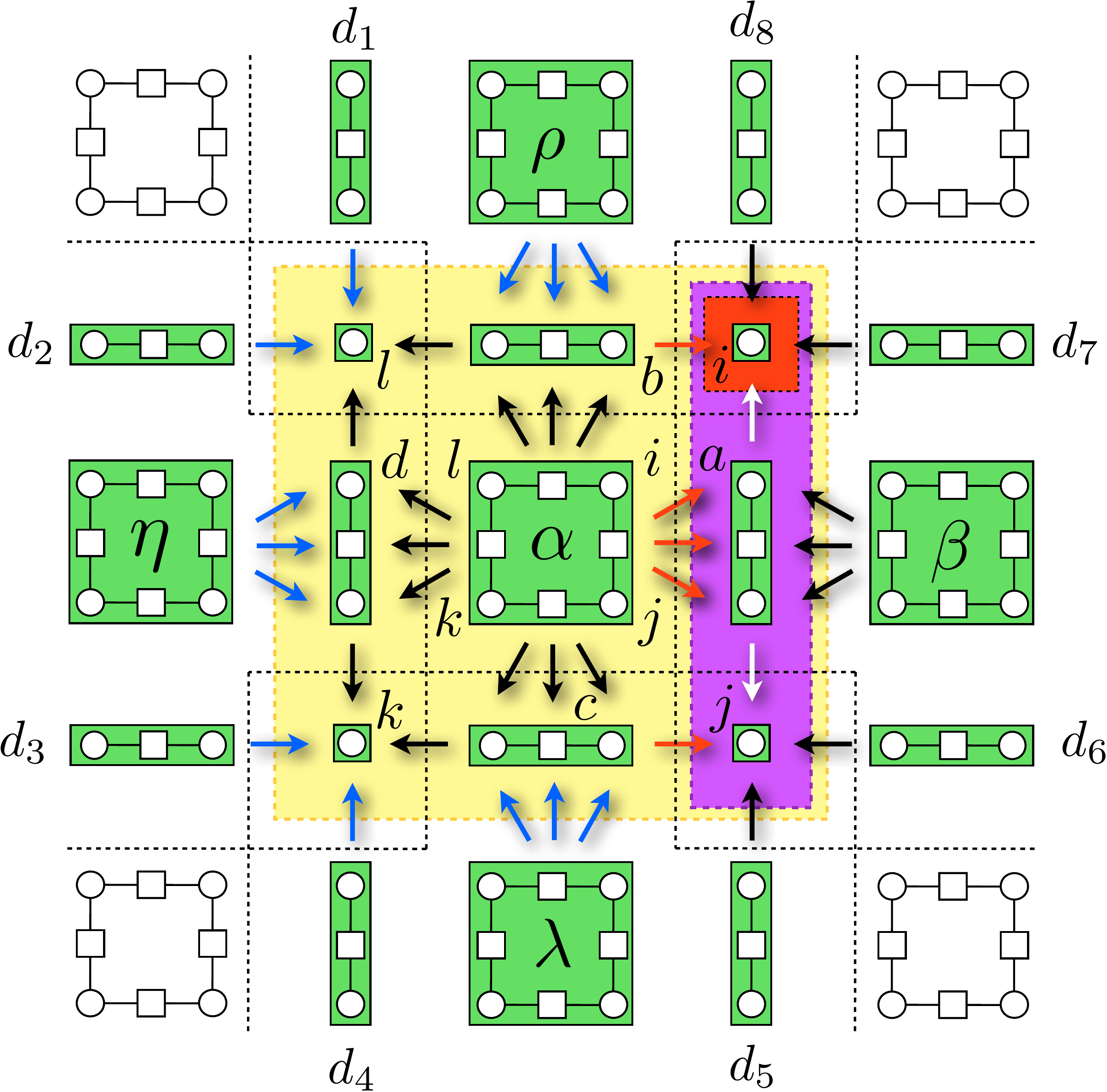}\hspace{1.5cm}
\includegraphics[width=6.2cm]{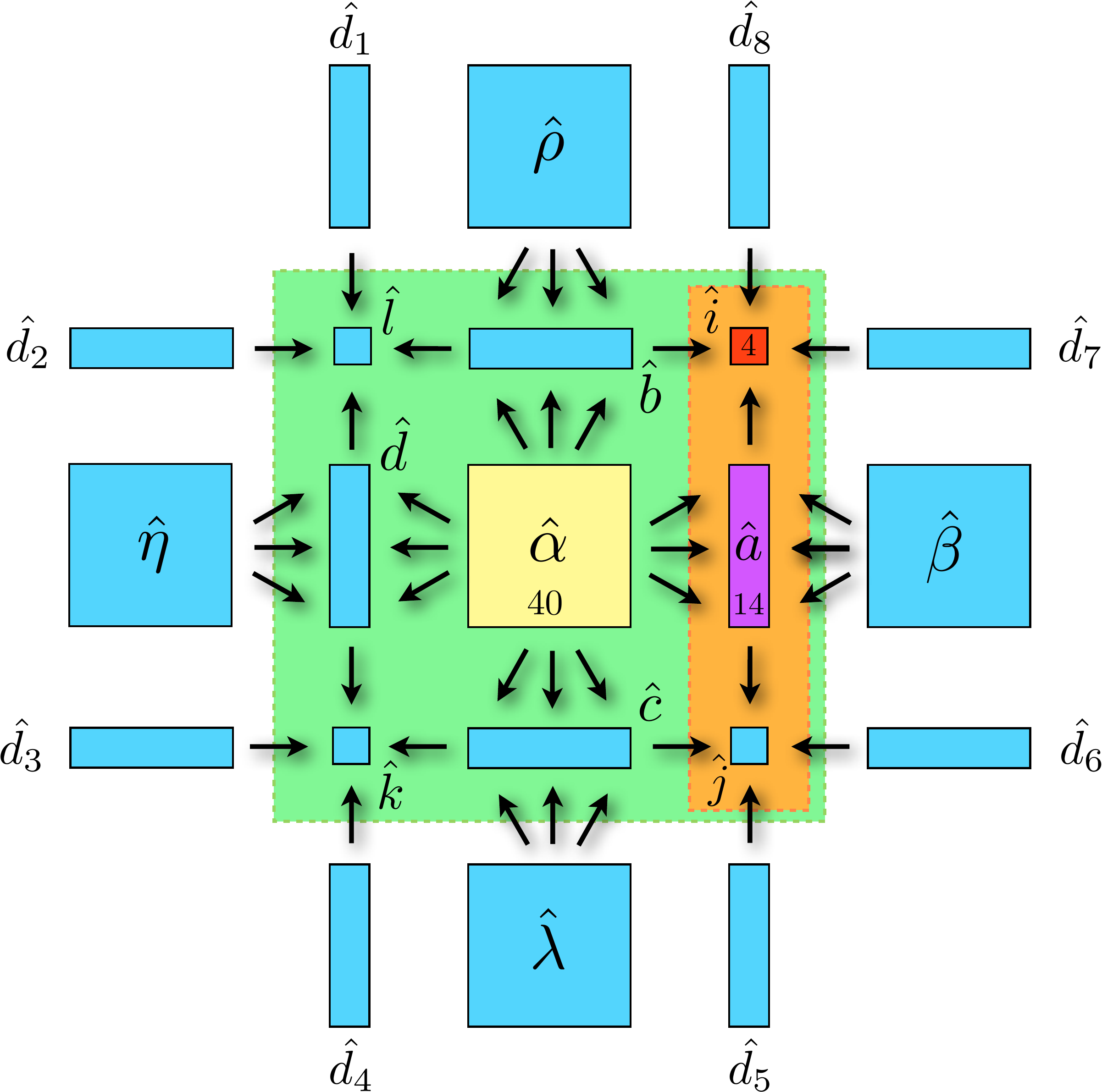}
\caption{\footnotesize Left panel: Region graph representation of a 2D EA model in terms of plaquettes, rods and vertices. The border and interior of the plaquette region $\alpha$ is pictured in yellow and becomes the region $\hat{\alpha}$ in the auxiliary model (yellow plaquette in the right panel). The border and interior of the rod region $a$ is pictured in purple and becomes the new rod region $\hat{a}$ (right panel) whereas the border of the vertex region $i$, which is displayed in red, becomes the new vertex region $\hat{i}$ in the auxiliary model. Right panel: the auxiliary graph in terms of new regions. Numbers inside the regions are a memo of the number of $\vec{U},\vec{u}$ field-messages contained in each region. The arrows represent the new messages in the auxiliary model, indicated as $Q(\vec U, \vec u)$ in the text.}
\label{fig:2}
\end{center}
\end{figure}

The second step is to define new regions in the auxiliary graph and assign the second-level potential functions and constraints to these 
regions. How to do so is not uniquely defined, no more than in the standard CVM.
We here choose a region graph for the auxiliary second-level graphical model which is
isomorphic to the region graph for the first-level graphical model. 
This means that plaquettes, rods and vertices on the first level will be in one-to-one correspondence with regions on the second level which we call by the same names.
To a plaquette $\alpha$ on the first-level graph corresponds a second-level region denoted by $\hat{\alpha}$
which contains \textit{(i)} all messages going to descendants of $\alpha$ from the outside or from the inside 
\textit{i.e.} $B_{\alpha} \bigcup I_{\alpha}$;
\textit{(ii)} the delta functions enforcing that the messages in  $I_{\alpha}$ satisfy GBP; 
\textit{(iii)} the potential function $Z_{\alpha}^y$. Similarly, to a rod $a$ on the first-level graph corresponds a second-level region denoted by $\hat{a}$ which contains all messages in $B_a \bigcup I_a$, the two delta functions enforcing that the messages in
$I_a$ satisfy GBP, and $Z_{a}^{\, y}$. To a vertex $i$ corresponds finally $\hat{i}$ which contains the four messages
in $B_i$, and $Z_{i}^{y}$.  
The procedure is illustrated in Fig \ref{fig:2}.

%
%
The third step is to carry out CVM on \eqref{eq:GPFfinal} with the regions defined in Fig.~\ref{fig:2} and define consistency conditions for marginals between parent-to-child regions similarly to those in \eqref{eq:consist} for the first-level region graph.
The second-level CVM is built on messages in the second-level region graph,
which are functions of the messages $\vec{m}$ or, rather, of the fields used to parametrize the messages in the first-level region graph and which we here denote as $Q(\vec U, \vec u)$ (probabilities of probabilities) illustrated by black arrows in Fig \ref{fig:2}, right panel. 
Every one of these (second-level) messages goes from a parent to a child region and depends on all the (first-level)
messages, \textit{i.e.} message-fields, in the child region. When there is a direct plaquette-to-rod link $\hat{\alpha} \to \hat{a}$ on the second-level region graph, the $Q_{\hat{\alpha} \to \hat{a}}$ message depends on all the fields in the region $\hat{a}$, that is all the fields in $B_a \bigcup I_a$ in the original graph and so depends on $14$ (first-level) field-messages. When there is a direct rod-to-vertex link $\hat{a} \to \hat{i}$ the $Q_{\hat{a} \to \hat{i}}$ message depends on all the fields in the region $\hat{i}$, corresponding to the region $B_i$ in the original graph and so includes $4$ fields (see Fig \ref{fig:2} for an illustration). 

The CVM and GBP construction at the first level introduce two types of $\vec m$ first-level messages, from plaquette-to-rod and from rod-to-vertex, which can be parametrized using $u$-fields as shown in \ref{Sec:GF}. Therefore on the second-level region, by a one-to-one correspondence with the first-level graph, there also exist two different kinds of $Q(\vec U, \vec u)$ messages. More explicitly, we represent these second-level messages in the auxiliary graph using a different notation for each one of them:
\begin{align}\label{eq:Q_plaq}
&Q_{\hat{p} \to \hat{r}}(\vec{U}_{p \to r }^{(2)}, \vec{u}_{p \to r}^{(v)\,(4)}, \vec{u}_{r \to v}^{(8)}) &&\quad \text{from plaquettes to rods,}\\
&q_{\hat{r} \to \hat{v}}( \vec{u}_{r \to v}^{(4)}) &&\quad \text{from rods to vertices,}
\label{eq:Q_rod}
\end{align}
where superscript numbers refers to how many messages of the kind $\mu \to \nu$ are contained in the argument of $Q$. 
That is, the notation $U_{p \to r}^{(l)}$ means $l$ field-messages of 
the kind plaquette-to-rod, $\vec{u}_{p \to r }^{(v)\,(l)}$ means $l$ field-messages of 
the kind plaquette-to-rod of the vertex type, whereas  $\vec{u}_{r \to v}^{(l)}$ means $l$ messages of the kind rod-to-vertex, according to notation introduced in Section \ref{Sec:GF}.
In the following, for the sake of clarity, we omit the labels and the numbers of the fields $\vec{u}$ when they are arguments of the functions $Q$ and $q$, \textit{i.e.} 
we simply write $Q_{\hat{p} \to \hat{r}}(\vec{U}, \vec{u})$ and $q_{\hat{r} \to \hat{v}}( \vec{u})$.

We can now look at the second-level region graph as a new auxiliary region graph on an higher level, which contains new messages (probabilities of probabilities), new potential functions and hard-core constraints. As for the first level-region graph (see eq. \eqref{eq:consist}), the new marginal probabilities, or beliefs, defined on the second-level region graph, which we here indicate as $W_{\hat{\alpha}}$, $W_{\hat{a}}$ and $W_{\hat{i}}$ for new plaquette, rod and vertex respectively, must satisfy probability consistency equation as   
\begin{align}\label{eq:consistP}
\int_{\hat{\alpha} \backslash \hat{a}} d \vec{U}^{(6)} d \vec{u}^{\,(20)} \, W_{\hat{\alpha}}(\vec{U}^{(8)},\vec{u}^{(32)}) &= W_{\hat{a}}(\vec{U}^{(2)},\vec{u}^{(12)})&&\quad \text{for plaquette $\hat \alpha$ and rod $\hat a$,}\\\label{eq:consistR}
\int_{\hat{a} \backslash \hat{i}} d \vec{U}^{(2)} d \vec{u}^{\,(8)} \, W_{\hat{a}}(\vec{U}^{(2)},\vec{u}^{(12)}) &= W_{\hat{i}}(\vec{u}^{(4)})&&\quad  \text{for rod $\hat a$ and vertex $\hat i$}.
\end{align}
Above the notation $\hat{\alpha} \backslash \hat{a}$ means that the integral is taken over all the messages belonging to $\hat{\alpha}$ except those which are common with $\hat{a}$ (see Fig. \ref{fig:gbp_arrows}, left panel) (and similarly for $\hat{a} \backslash \hat{i}$, see Fig. \ref{fig:gbp_arrows}, central panel). Analogously to the marginals \eqref{eq:omega} in the first-level region graph, the second-level beliefs are defined through the new messages $Q$ and $q$, new potential functions $Z$ and hard-constraints included in each second-level region. Since equations \eqref{eq:consistP} and \eqref{eq:consistR} are the core of our approach to second-level GBP we will write them down explicitly in the following two subsections.


\subsection{Consistency equation for rods and vertices}\label{Sec:Cons_RV}

In analogy with the consistency equations \eqref{eq:consist}, we here write explicitly the consistency equations for the second-level region graph illustrated in Fig. \ref{fig:2}. 
As mentioned, we interpret the second-level graph as a new auxiliary graph having new potential functions. These are given by the partition functions $Z$ re-weighted by the Parisi parameter $y$, new messages between parent-to-child regions given by the probabilities of probabilities $Q$ and $q$ and, in addition to the original region graph, contains new constraints which enforce the new variables $\vec U$ and $\vec u$ of the second-level graph to satisfy the GBP conditions on the first-level region graph. 

The new marginals can then be deduced by analogy with the first-level marginals. The $Q$ and $q$ messages involved in the marginal of a region are all those from parent-to-child region $\hat \mu \to \hat \nu$ where $\hat \mu$ belongs to the border and $\hat \nu$ belongs to the interior of the child region. The potential function for each second-level region must count for all the energetic interactions of that region and therefore it is given by the partition function of the same region which contains, indeed, all the energetic terms. Finally the hard-core constraints should enforce all the GBP conditions for the messages internal to the region and therefore they have to be all the delta functions which belong to the interior of the region considered. 

With this procedure as a tool, using a compact notation, the local consistency equation \eqref{eq:consistR} can be written as follows: 
\begin{align} 
\frac{1}{\Xi_{\hat{a}}} \int_{\hat{a} \backslash \hat{i}} d \vec{U} d \vec{u}\,  Z_{a}^{\,^{y}}(\vec U, \vec u) \,\delta_{a \to i}^{}\,\delta_{a \to j}^{} \prod^{(2)}Q_{\hat{\mu} \to \hat{\nu}}(\vec{U}, \vec u) \,\prod^{(6)} q_{\hat{\mu} \to \hat{\nu}}(\vec u) = \frac{1}{\Xi_{\hat{i}}} Z_{i}^{^{y}}(\vec u) \, \prod^{(4)} q_{\hat{b} \to \hat{i}}(\vec{u})
\label{eq:cons_rod_vert}
\end{align}
where ${\Xi_{\hat{a}}}$ and ${\Xi_{\hat{i}}}$ are normalization factors and the superscript on the products indicate how many $Q$ or $q$ products we have on each side.
The delta functions in the equation above enforce the GBP equations \eqref{eq:field_LV}  for rods-to-vertex messages and are interpreted as hard-core constraints; for short we use the notation $\delta_{\mu \to \nu} = \delta{[u_{\mu \to \nu}-\mathcal{L}_{\mu \to \nu}(\hat{\ul U}, \hat{\ul u})]}$, where the function $\mathcal{L}_{\mu \to \nu}$ is defined in \eqref{eq:field_LV} or explicitly in \eqref{eq:RtoV_field}. The messages constrained by these two deltas are pictorially represented as white arrows in the purple rod region of Figure \ref{fig:2}, left panel. 
The $Z$'s are potential functions referred to the region considered and the Parisi parameter $y$ plays the role of an inverse temperature. The different kind of $Q$'s on the LHS are all those from the parent-to-child region $\hat \mu \to \hat \nu$ with $\hat \mu \in B(\hat a)$ and $\hat \nu \in I(\hat a)$ and the products are on these same variables. Pictorially, they are represented by all the arrows crossing the perimeter of the orange rod region of Figure \ref{fig:2}, right panel. Analogously, all the $Q$'s and the products on the RHS are all those from $\hat \mu \to \hat i$ where $\hat \mu \in B(\hat i)$ therefore, pictorially, the four black arrows pointing towards the red region $\hat i$ in the same figure.

Similarly to what pointed out for equations \eqref{eq:noJacobian} and \eqref{eq:Jacobian}, the introduction of delta functions with functional dependence on the GBP equations above would, in principle, involve a determinant on the LHS of \eqref{eq:cons_rod_vert}. This determinant is of the type $| \textit{\rm{det}}(\mathds{1}-\Delta \mathcal{L})| $ where $\Delta \mathcal{L}$ is the matrix containing the derivatives of the GBP rod-to-vertex function $\mathcal{L}(\vec U, \vec u)$, respect to all the fields $\vec U$ and $\vec u$ belonging to the rod region $\hat a$, computed at the fixed point. As shown in Appendix \ref{s:determinants}, it turns out that this determinant is equal to one and therefore brings no correction to the above equation. 

Writing more explicitly \eqref{eq:cons_rod_vert} for the rod $\hat{a}$ and the vertex $\hat{i}$ on the expanded graph, with reference to Fig. \ref{fig:2} for labelling, we have
\begin{align} \nm
\frac{1}{\Xi_{\hat{a}}} \int_{\hat{a} \backslash \hat{i}} & d \vec{U}^{} d \vec{u}^{}  \, Z_{a}^{\,^{y}}(\vec U, \vec u) \,\delta_{a \to i}^{}\,\delta_{a \to j}^{} \,Q_{\hat{\alpha} \to \hat{a}}(\vec U, \vec u)  Q_{\hat{\beta} \to \hat{a}}(\vec U, \vec u) \,\dashuline{q_{\hat{b} \to \hat{i}}(\vec u) \, q_{\hat{d}_8 \to \hat{i}}(\vec u) \, q_{\hat{d}_7 \to \hat{i}}(\vec u)}\, \\
&\times q_{\hat{c} \to \hat{j}}(\vec u) \, q_{\hat{d}_5 \to \hat{j}}(\vec u) \, q_{\hat{d}_6 \to \hat{j}}(\vec u)= \frac{1}{\Xi_{\hat{i}}} Z_{i}^{^{y}}(\vec u) \, q_{\hat{a} \to \hat{i}}(\vec u)\, \dashuline{q_{\hat{b} \to \hat{i}}(\vec u)\, q_{\hat{d}_8 \to \hat{i}}(\vec u)\, q_{\hat{d}_7 \to \hat{i}}(\vec u)}
 \label{eq:GBP_rod_vertex}
\end{align}
where the common terms on the left and right-hand-side have been underlined by a dashed line. Let us observe that the integration does not involve these underlined terms. Indeed these terms only depend  on the $\vec u$ field-messages which belong to the region $\hat i$ whereas the integration is over the $\vec u$ in the set $\hat a \backslash \hat i$. Referring to Figure \ref{fig:gbp_arrows}, central panel, where the $\vec u$ are illustrated by arrows, the integration only involves the blue $\vec u$'s in the purple region $\hat a$ and does not include the black fields shared with the red region $\hat i$.
Therefore since the underlined $q$'s never depend on the variable of integration they can be canceled on both sides of the equation. The result reads
\begin{align}
q_{\hat{a} \to \hat{i}}(\vec u)=&\frac{\Xi_{\hat{i}}}{\Xi_{\hat{a}}} \int_{\hat{a} \backslash \hat{i}} d \vec{U}^{} d \vec{u}^{}  \, Z_{a\to i}^{\,^{y}}(\vec U, \vec u) \,\delta_{a \to i}^{}\,\delta_{a \to j}^{} Q_{\hat{\alpha} \to \hat{a}} \, Q_{\hat{\beta} \to \hat{a}} \, q_{\hat{c} \to \hat{j}} \, q_{\hat{d}_5 \to \hat{j}} \, q_{\hat{d}_6 \to \hat{j}}
\label{eq:Q2}
\end{align}
where we used the relation \eqref{eq:Zai}, \textit{i.e.} $Z_a= Z_{a \to i}\int Z_i \, \delta_{a \to i}$ expressed in terms of field-messages, and  we omitted the field dependencies of the $Q$ and $q$ second-level messages for brevity. 
We observe that the field $u_{a \to j}$ is integrated over on RHS of \eqref{eq:Q2} and, in addition,  $Z_{a \to i}$ does not depend on this message (see \eqref{eq:Zai} for details). We then integrate over this field and make use of the delta function $\delta_{a \to j}$, we get
\begin{align}
q_{\hat{a} \to \hat{i}}(\vec u)=&\frac{\Xi_{\hat{i}}}{\Xi_{\hat{a}}} \int_{\hat{a} \backslash \hat{i}} d \vec{U}^{} d \vec{u}^{}  \, Z_{a\to i}^{\,^{y}}(\vec U, \vec u) \,\delta_{a \to i}^{}\,\tilde Q_{\hat{\alpha} \to \hat{a}} \, \tilde Q_{\hat{\beta} \to \hat{a}} \, \tilde q_{\hat{c} \to \hat{j}} \, \tilde q_{\hat{d}_5 \to \hat{j}} \, \tilde q_{\hat{d}_6 \to \hat{j}}
\label{eq:Qrod1}\, .
\end{align}
With the tilde notation above we indicated functions $Q$ and $q$ with the value of $u_{a \to j}$ in their argument replaced by their GBP updates \eqref{eq:field_LV} \textit{i.e.} by $\mathcal{L}_{a \to j}(\hat{\ul{U}}, \hat{\ul{u}})$. Equation \eqref{eq:Qrod1} corresponds to the 1RSB rod-to-vertex Generalized Belief Propagation equation for a 2D lattice model as the EA model and represent the first result of this paper. We recall that the $\Q$ and $q$ functions above, as their tilde version, are joint probability distributions of several messages therefore equation \eqref{eq:Qrod1}, although analytically consistent, it is very hard, perhaps impossible, to iterate numerically. In Section \ref{s:no-GSP} we discuss how to take an SP-like ansatz on the $\Q$ and $q$ functions to simplify the above equation. 


\subsection{Consistency equation for plaquettes and rods}\label{sec:ConsEqPR}

Similarly to the previous section and with an analogous notation, we here want to write more explicitly the consistency equations between plaquettes and rods on the second-level region graph. Equation \eqref{eq:consistP} can be written as 
\begin{align} \nm
\frac{1}{\Xi_{\hat{\alpha}}} &\int_{\hat{\alpha} \backslash \hat{a}}  d \vec{U} d \vec{u}\,  Z_{\alpha}^{\,^{y}}(\vec U, \vec u) \, \delta_{p \to r}^{(4)}\, \delta_{p \to r}^{(v)\, (8)}\,\delta_{r \to v}^{(8)} \prod^{(4)} Q_{\hat{p} \to \hat{r}}(\vec U, \vec u) \prod^{(8)} q_{\,\hat{r} \to \hat{v}}(\vec u) \\
&\hspace{1cm}= \frac{1}{\Xi_{\hat{a}}} Z_{a}^{^{y}}(\vec U, \vec u) \,\delta_{a \to i}\, \delta_{a \to j} \prod^{(2)} Q_{\hat{p} \to \hat{r}}(\vec U, \vec u) \prod^{(6)} q_{\,\hat{r} \to \hat{v}}(\vec u)
\label{eq:GBP2-plaquette-to-rod}
\end{align}
where $\Xi_{\hat{\alpha}}$ and $\Xi_{\hat{a}}$ are normalization factors. The second-level messages which appear on the LHS are all those from the parent-to-child region $\hat \mu \to \nu$ with $\hat \mu \in B(\hat \alpha)$ and $\hat \nu \in I(\hat\alpha)$ and the products are over these variables. Pictorially they are represented in Figure \ref{fig:2}, right panel, by all the arrows going through the perimeter of the square green region. The $Q$'s and $q$'s on the RHS are those going from $\hat \mu \to \hat \nu$ with $\hat \mu \in B(\hat a)$ and $\hat \nu \in I(\hat a)$ and are illustrated, in the same aforementioned picture, as the arrows crossing the perimeter of the rod orange region. The product on this side of the equations are on these same variables. 

Delta functions in the equation above enforce GBP equations for the field-messages at both the plaquette-to-rod and rod-to-vertex level and, as for \eqref{eq:cons_rod_vert}, are interpreted as hard-core constraints for the second-level region graph model. Their upper script refers to how many delta functions of the type $\mu \to \nu$ are included. All those on the LHS of \eqref{eq:GBP2-plaquette-to-rod} enforce constraints for the field-messages of the region $I(\alpha)$ whereas those on the RHS enforce constraints for the region $I(a)$. Note that since $I(a) \subset I(\alpha)$ some constraints appear on both sides. 

More explicitly, on the LHS four deltas enforce GBP plauqette-to-rod equations \eqref{eq:fields_PL1} for the first level field-messages $U_{p \to r}$ as $\delta_{p \to r} = \delta{[U_{p \to r}-\mathcal{F}_{p \to r}(\hat{\ul U}, \hat{\ul u})]}$ and eight deltas enforce equations \eqref{eq:fields_PL2} and \eqref{eq:fields_PL3}  for the field messages $u_{p \to r}^{(v)}$ as $\delta_{p \to r}^{(v)} = \delta{[u_{p \to r}^{(v)}-\mathcal{G}_{p \to r}^{(v)}(\hat{\ul U}, \hat{\ul u})]}$. The remaining eight deltas enforce the GBP link-to-vertex conditions \eqref{eq:field_LV} as $\delta_{r \to v} = \delta{[u_{r \to v}-\mathcal{L}_{r \to v}(\hat{\ul U}, \hat{\ul u})]}$ and two of them appear also on the RHS. For clarity, referring to Figure \ref{fig:2}, left panel,  the field messages $U$ and $u$ constrained by the delta functions on the LHS of \eqref{eq:GBP2-plaquette-to-rod}  are illustrated as black, red and white arrows in the interior of the yellow square region $\hat \alpha$. Whereas the field-messages on the RHS of the same equation constrained by the two delta conditions are illustrated as white arrows in the purple rod region $\hat{a}$ of the same figure.

 Note that, as for the rod-to-vertex equation, the introduction of the delta functions would, in principle, involve a determinant term on both sides. The LHS of \eqref{eq:GBP2-plaquette-to-rod} should be multiplied by $| \textit{\rm{det}}( \mathds{1}-\Delta \mathcal{F})|$
where $\Delta \mathcal{F}$ is the matrix of all the derivatives of the GBP plaquette-to-rod and rod-to-vertex equations respect to all the field-messages in the plaquette. Similarly, the RHS should be multiplied by $| \textit{\rm{det}}( \mathds{1}-\Delta \mathcal{L})|$ which is the same determinant, equal to one, encountered in Sec. \ref{Sec:Cons_RV}. It turns out that, after fixing the gauge of the GBP equations as shown by the rest of this section, also $| \textit{\rm{det}}( \mathds{1}-\Delta \mathcal{F})|$ is equal to one (see Appendix \ref{s:determinants}) and therefore there exists no correction to the one-step replica symmetric GBP equations because of the determinant, neither for the plaquette-to-rod nor for the rod-to-vertex equations. 

Writing equation \eqref{eq:GBP2-plaquette-to-rod} more explicitly, with labeling referred to Figure \ref{fig:2} we get:
\begin{align}\label{eq:LC} \nm	
\frac{1}{\Xi_{\hat{\alpha}}}  \int_{\hat{\alpha} \backslash \hat{a}} d \vec{U} d \vec{u}\, & Z_{\alpha}^{\,^{y}}(\vec U, \vec u) \,\dashuline{ Q_{\beta \to a} } Q_{\rho \to b}  \,Q_{\eta \to d} \, Q_{\lambda \to a} \,q_{d_1 \to l} q_{d_2 \to l} q_{d_3 \to k} q_{d_4 \to k} \dashuline{q_{d_5 \to j} q_{d_6 \to j} q_{d_7 \to i} q_{d_8 \to i}}\\ \nm
&\times\delta_{\alpha \to a}^{(3)}\, \delta_{\alpha \to b}^{(3)} \,\delta_{\alpha \to c}^{(3)}\, \delta_{\alpha \to d}^{(3)} \,\delta_{a \to i}\,\delta_{a \to j}\,\delta_{c \to j}\,\delta_{c \to k}\delta_{d \to k}\,\delta_{d \to l}\,\delta_{b \to l}\,\delta_{b \to i}\,\\ 
&=\frac{1}{\Xi_{\hat{a}}} Z_{a}^{^{y}}(\vec U, \vec u) \,\delta_{a \to i}\, \delta_{a \to j} \dashuline{Q_{\beta \to a}}\, Q_{\alpha \to a} q_{c \to j} \dashuline{q _{d_5 \to j} q_{d_6 \to j} q_{d_7 \to i} q_{d_8 \to i}} q_{b \to i}
\end{align}
where, on both side of the equation, common probability terms have been underlined and field dependencies of the $Q$ and $q$ second-level messages are omitted. To shorten notation, terms like $\delta_{\alpha \to a}^{(3)}$ above stands for $\delta_{\alpha \to a}\,\delta_{\alpha \to a}^{(i)}\, \delta_{\alpha \to a}^{(j)}$ and enforce GBP constraints for the fields $U_{\alpha \to a}, u_{\alpha \to a}^{(i)}$ and $u_{\alpha \to a}^{(j)}$ respectively.  Let us note that all the underlined probabilities of field-messages have no dependence on the fields which belong 
to the region $\hat \alpha \backslash \hat a$ but, rather, they only depend on the fields fields $\vec U$ and $\vec u$ which are in the second-level region $\hat a$. Using the left panel of Figure \ref{fig:gbp_arrows} as a reference, the underlined second level messages only depend on the field-messages illustrated as black arrows which cross or are internal of the purple rod region $\hat a$. Therefore since the integration on the LHS of \eqref{eq:LC} is only over the field-messages in $\hat \alpha \backslash \hat a$ (the blue arrows in the aforementioned figure), the underlined $Q$'s and $q$'s can be simplified on both sides. Using the relation among partition functions of different regions contained in the Appendix, \textit{i.e.} \eqref{eq:Zalphaa} here written in terms of field messages as $Z_\alpha = Z_{\alpha \to a} \int Z_a \,\delta_{\alpha \to a}\,\delta_{\alpha \to a}^{(i)}\, \delta_{\alpha \to a}^{(j)}$, we can then rewrite \eqref{eq:LC} as 
\begin{align}\label{eq:P-L_simp} \nm	
\delta_{a \to i}\, \delta_{a \to j}\, Q_{\alpha \to a} \,q_{b \to i}\, q_{c \to j} =& \frac{\Xi_{\hat{a}}}{\Xi_{\hat{\alpha}}}  \int_{\hat{\alpha} \backslash \hat{a}} d \vec{U} d \vec{u}\,  Z_{\alpha \to a}^{\,^{y}}(\vec U, \vec u) \, Q_{\rho \to b} \,Q_{\eta \to d} \, Q_{\lambda \to a} q_{d_1 \to l} q_{d_2 \to l} q_{d_3 \to k} q_{d_4 \to k} \\ \nm
&\times\delta_{\alpha \to a}\delta_{\alpha \to a}^{(i)} \delta_{\alpha \to a}^{(j)}\, \delta_{\alpha \to b}\delta_{\alpha \to b}^{(l)} \delta_{\alpha \to b}^{(i)} \,\delta_{\alpha \to c}\delta_{\alpha \to c}^{(j)} \delta_{\alpha \to c}^{(k)}\, \delta_{\alpha \to d}\delta_{\alpha \to d}^{(k)} \delta_{\alpha \to d}^{(l)} \\
&\times\delta_{a \to i}\,\delta_{a \to j}\,\delta_{c \to j}\,\delta_{c \to k}\,\delta_{d \to k}\,\delta_{d \to l}\,\delta_{b \to l}\,\delta_{b \to i}
\end{align}
Let us note that the two deltas $ \delta_{a \to i}[u_{a \to i}- \mathcal{L}_{a \to i}(\hat{\ul{U}}, \hat{\ul{u}})]$ and $ \delta_{a \to j}[u_{a \to j}- \mathcal{L}_{a \to j}(\hat{\ul{U}}, \hat{\ul{u}})]$ only depend on field-messages in $\hat{a}$, \textit{i.e.} on first-level field-messages which are not integrated over in 
the above equation. Therefore, although $\hat{a}$ contains fourteen field-messages,  
equation \eqref{eq:P-L_simp} only holds information for the subset where the two messages 
$u_{a\to i}$ and $u_{a\to j}$ in $I(a)$ are determined by the eight messages in $B(a)$ through the constraints  $\delta_{a \to i}$ and 
$\delta_{a \to j}$. Otherwise the meaning of this aforementioned equation is trivially $0=0$. 

In the following we want to proceed integrating out all the deltas on the RHS of \eqref{eq:P-L_simp} which do not enforce constraints on those fields-messages directed to the region $I(a)$. Referring to Figure \ref{fig:gbp_arrows}, left panel, we want to integrate out all the fields internal to the yellow square region $I(\alpha)$ (pictured as blue arrows), except those directed to or internal at the purple regions $I(a)$ (pictured as black arrows). 

First of all we impose gauge fixing conditions for the GBP equations at the plaquette level as discussed in Sec. \ref{Sec:GF} and in reference \cite{dominguez2011characterizing}. According to this gauge fixing procedure, at the first-level region graph we set to zero four of the messages $u_{\alpha \to r}^{(v)}$ in $I(r)$. This means that, at the second-level region graph, four over eight deltas $\delta_{\alpha \to r}^{(v)}(u_{\alpha \to r}^{(v)}-G_{\alpha\to r}^{(v)}(\vec U, \vec u))$ are changed to $\delta(u_{\alpha \to r}^{(v)})$. For each rod in the region graph there exist two of these deltas (see for example Fig \ref{fig:gbp_arrows}), the choice of which to change accordingly to the fixing gauge conditions discussed in \cite{dominguez2011characterizing} is totally arbitrary, 
we then choose to change  $\delta_{\alpha \to a}^{(i)}, \delta_{\alpha \to b}^{(l)}, \delta_{\alpha \to d}^{(k)}, \delta_{\alpha \to c}^{(j)}$. Among these deltas, we integrate out those not referred to messages heading to $a$, therefore just $\delta_{\alpha \to b}^{(l)}, \delta_{\alpha \to d}^{(k)}, \delta_{\alpha \to c}^{(j)}$ . The effect of this integration on the rest of the remaining terms is to set to zero the related field-messages wherever else they appear. 

We then integrate out three over four of the delta functions which fix the GBP conditions for the fields-messages $U_{\alpha \to r}$, \textit{i.e.} which enforce \eqref{eq:fields_PL1}, in particular we integrate out $\delta_{\alpha \to b}, \delta_{\alpha \to c}, \delta_{\alpha \to d}$. Let us observe that the function $\mathcal{F}(\vec U, \vec u)$ which appears in $\delta_{\alpha \to r}(U_{\alpha \to r}- \mathcal{F}_{\alpha \to r}(\vec U, \vec u))$ only depends on the external field-messages in $B(\alpha)$ which, at the GBP level for the plaquette $\alpha$, are known values. Therefore the effect of this integration on the remaining terms is to change the value of the $U_{\alpha \to r}$ by the values of some external field-messages, wherever they appear. 

We continue by integrating out three over four of the remaining $u_{\alpha \to r}^{(v)}$ field-messages from plaquette-to-rod of the vertex type and make use of the relative delta functions which enforce the GBP equations \eqref{eq:fields_PL3} for these fields.
Explicitly we integrate out $\delta_{\alpha \to b}^{(i)}, \delta_{\alpha \to d}^{(l)}, \delta_{\alpha \to c}^{(k)}$. Let us note that the RHS of the equations \eqref{eq:fields_PL3} enforced by these deltas is a function of the external messages, \textit{i.e.} messages in $B(\alpha)$, and depend linearly on one message which belong to $I(\alpha)$. To write it more explicitly, we could for instance write one of these deltas as $\delta_{\alpha \to a}^{(j)}(u_{\alpha \to a}^{(j)} - (h(\vec U_{ex}, \vec u_{ex})-u_{c \to j}))$ where the label $ex$ stands for ``externals'', \textit{i.e.} those fields are in $B(\alpha)$ and $h(\vec U_{ex}, \vec u_{ex})= \mathcal{G}(\vec U, \vec u) + u_{c \to j}$ from equation \eqref{eq:fields_PL3}. Therefore, the integration over these deltas has the effect of changing messages of the kind $u_{\alpha \to a}^{(j)}$ into a function of the external messages and of one internal message of the kind $u_{c \to j}$. 

We observe that this effect reflects on the link-to-vertex delta function $\delta_{b \to l}, \delta_{d \to k}, \delta_{c \to j}$ in a peculiar way. Indeed after these and previous integrations, the GBP functions contained in these deltas become only functions of the external messages, for instance $\delta_{c \to j}(u_{c \to j} - \mathcal{L}_{c \to j}(\vec U_{ex}, \vec u_{ex}))$. This is due to the linear dependence that these latter delta functions have on the internal message of the kind $u_{c \to i}$ and it entails simplifications (see equation \eqref{eq:RtoV_field} for details). 

We carry on by integrating out $\delta_{b \to l}, \delta_{d \to k}$ which only bring dependencies on the external messages in the remaining terms. After these latter integration, we are left with the deltas for the fields both heading to and internal at the region $I(a)$, \textit{i.e.} $\delta_{a \to j}, \delta_{a \to i}, \delta_{\alpha \to a}, \delta_{\alpha \to a}^{(i)}, \delta_{\alpha \to a}^{(j)}, \delta_{b \to i}, \delta_{c \to j}$ and, in addition, with $\delta_{d \to l}(u_{d \to l} - \mathcal{L}_{d \to l}(\vec U_{ex}, \vec u_{ex}, u_{c \to k}))$ and $\delta_{c \to k}(u_{c \to k} - \mathcal{L}_{c \to k}(\vec U_{ex}, \vec u_{ex}, u_{a \to j}))$. Therefore if we now integrate out $\delta_{d\to l}$,  the field $u_{d\to l}$ will be replaced by a function of both the external messages and $u_{c\to k}$. We carry on by integrating out also $\delta_{c\to k}$, as a consequence $u_{c\to k}$ will be replaced by a function of both the external messages and of $u_{a \to j}$. Equation \eqref{eq:P-L_simp} after all these integrations reads as:
\begin{align} \label{eq:Q1}\nm	
\delta_{a \to i}\, \delta_{a \to j}\, Q_{\alpha \to a} \, q_{b \to i} \, q_{c \to j} =& \frac{\Xi_{\hat{a}}}{\Xi_{\hat{\alpha}}}  \int_{\hat{\alpha} \backslash \hat{a}} d \vec{U} d \vec{u}\,  Z_{\alpha \to a}^{\,^{y}}(\vec U, \vec u) \, \tilde Q_{\rho \to b} \,\tilde Q_{\eta \to d}\, \tilde Q_{\lambda \to c} \,\tilde q_{d_1 \to l}\, \tilde q_{d_2 \to l}\, \tilde q_{d_3 \to k} \,\tilde q_{d_4 \to k} \\
&\times\delta_{\alpha \to a}\,\delta_{\alpha \to a}^{(i)}\, \delta_{\alpha \to a}^{(j)}\,\tilde  \delta_{b \to i}\,\tilde \delta_{c \to j}\,\delta_{a \to i}\,\delta_{a \to j}
\end{align}
where the tilde notation of the second-level messages and on the deltas means that some of their arguments have been replaced by their GBP updates because of the integration over all the other deltas listed above. Let us observe that both sides of the equation \eqref{eq:Q1} depend on all the field-messages in the region $\hat a$, \textit{i.e.} $B(a) \bigcup I(a)$ which are pictorially represented in Figure \ref{fig:gbp_arrows}, left panel, as all the black arrows entering and internal to the purple rod region. Equation \eqref{eq:Q1} corresponds to the 1RSB plaquette-to-rod GBP equation for the 2D EA model. Since the $\Q$'s and $q$'s functions above are joint probability distribution of messages, similar considerations as those in the text below \eqref{eq:Qrod1} apply here. 


\section{A class of generalized Survey Propagation equations}
\label{s:no-GSP}
We start by explaining the approach to Survey Propagation (SP) described in \cite{mezard2009information}.
The success of Belief Propagation is to a large extent due to its moving information forward. Indeed, this is why these algorithms are referred to as propagation. SP is a special class of solutions to the general second-level, 1RSB, BP equations which has the same property and for which the fixed points can therefore be found by forward iteration.

In the following we  want to verify whether the second-level 1RSB GBP equations presented in the previous section admit a special class of solutions of the SP-type and can therefore be written as a class of Generalized Survey Propagation equations.  The special solutions of SP are derived by assuming that a second-level message $q_{\hat{r}\to\hat{v}}$ only 
depends on the one first-level message going the same way, \textit{i.e.} on  $m_{a\to i}$ \cite{mezard2009information}.
Taking the same assumption, in our case, means that all the $\Q_{\hat p \to \hat r}$'s only depend on the underlining message $m_{\hat p \to \hat r}$ function of two spins and therefore parametrized by three $u$-fields, whereas all the $q_{\hat r  \to \hat v}$'s depend on the message $m_{\hat r  \to \hat v}$ function of one spin and therefore parametrized by one $u$-field. In other words, plaquette-to-rod messages as $Q_{\hat \alpha \to \hat a}$ will be assumed to be function like $Q_{\hat \alpha \to \hat a}(U_{\alpha \to a}, u_{\alpha \to a}^{(i)}, u_{\alpha \to a}^{(j)})$ and rod-to-vertex message to be $q_{a \to i}(u_{a \to i})$.
Following the SP ansatz we assume that this is so for all the second-level messages on RHS of the rod-to-vertex equation \eqref{eq:Q2} and we verify that the assumption is preserved on the LHS. We observe that, the field-message $u_{a\to j}$ only appears on RHS as an argument of the delta function $\delta_{a\to j}$ which can be integrated out to give 
\begin{align}\nm
q_{\hat{a} \to \hat{i}}(u_{a \to i}, u_{b \to i}, & u_{d_7 \to i}, u_{d_8 \to i})\\ \nm
= &\frac{\Xi_{\hat{i}}}{\Xi_{\hat{a}}} \int_{\hat{a} \backslash \hat{i}} d \vec{U}^{} d \vec{u}^{}  \, Z_{a\to i}^{\,^{y}}(U_{\alpha \to a}, u_{\alpha \to a}^{(i)}, u_{\alpha \to a}^{(j)} U_{\beta \to a}, u_{\beta \to a}^{(i)}, u_{\beta \to a}^{(j)}, u_{c\to j},u_{d_5\to j}, u_{d_6\to j}) \\ 
&\times\delta_{a \to i}^{}\, Q_{\hat{\alpha} \to \hat{a}} \, Q_{\hat{\beta} \to \hat{a}} \, q_{\hat{c} \to \hat{j}} \, q_{\hat{d}_5 \to \hat{j}} \, q_{\hat{d}_6 \to \hat{j}}.
\label{eq:Qrod2}
\end{align}
Above we have written out the arguments of $q_{\hat{a} \to \hat{i}}$ (on LHS) and
$Z_{a \to i}$ (on RHS) explicitly and we recall that the integration turns out to be over all the messages appearing as argument of $Z_{a \to i}$. Compared to \eqref{eq:Qrod1} we have now gained that the $Q$ and $q$ functions appearing on RHS are the original ones which, by assumption, still only depend on one underlying first-level message.
Let us now consider on what arguments actually depends $q_{\hat{a} \to \hat{i}}$. Obviously it depends on its first argument
since $u_{a\to i}$ appears explicitly on RHS in the delta function $\delta_{a \to i}(u_{a\to i}-\mathcal{L}_{a \to i}(\vec U, \vec u))$. However, it does not depend on its other three arguments since neither $u_{b\to i}$ nor $u_{d_7\to i}$ nor $u_{d_8\to i}$ appear on RHS. Therefore the assumption that second-level messages 
only depend on the underlying first-level message going the same way is preserved by the iteration of \eqref{eq:Qrod2}. We observe, in addition, that on a tree topology the above equation reduces to the SP equation for the single instance case \cite{mezard2009information,MezardParisiZecchina2002}. 

We then consider the plaquette-to-rod equation \eqref{eq:Q1} and take the same kind of SP-ansatz for the second-level messages on its RHS and observe that, as for equation \eqref{eq:Qrod2}, they no longer depend on several external messages brought into their argument by the integration of the internal delta functions performed in the previous section. We therefore remove the tilde notation for them and, after the SP-like ansatz, equation  \eqref{eq:Q1} reads:
\begin{align}\nm	
\delta_{a \to i}\, \delta_{a \to j}\, Q_{\alpha \to a}\, q_{b \to i} \, q_{c \to j} =& \frac{\Xi_{\hat{a}}}{\Xi_{\hat{\alpha}}}  \int_{\hat{\alpha} \backslash \hat{a}} d \vec{U} d \vec{u}\,  Z_{\alpha \to a}^{\,^{y}}(\vec U, \vec u) \, Q_{\rho \to b} \, Q_{\eta \to d} Q_{\lambda \to a}    q_{d_1 \to l}  q_{d_2 \to l}  q_{d_3 \to k}  q_{d_4 \to k} \\ 
&\times\delta_{\alpha \to a}\delta_{\alpha \to a}^{(i)} \delta_{\alpha \to a}^{(j)} \tilde  \delta_{b \to i}\,\tilde \delta_{c \to j}\,\delta_{a \to i}\,\delta_{a \to j}
\label{eq:Qsp}
\end{align}
To shorten notation we here do not write explicitly all the
dependencies of the partition function $Z_{\alpha \to a}$ but it turns
out that it depends exactly on all the $U$ and $u$ messages which
appear as argument of the SP-like $Q$'s and $q$'s on the RHS (see
Appendix \ref{ref:useful-relations}) and, as for \eqref{eq:Qrod2}, the
integration is precisely on all of them. Using the left panel of
Figure \ref{fig:gbp_arrows} as a reference, the integration on the RHS is over
all the field-messages represented as blue arrows entering the yellow
square region. Let us now look at which are the field dependencies of
the RHS of \eqref{eq:Qsp}. It depends on the fields appearing in the
delta functions $\delta_{\alpha \to a}, \delta_{\alpha \to a}^{(i)},
\delta_{\alpha \to a}^{(j)},\tilde \delta_{b \to i}$ and $\tilde
\delta_{c \to j}$ which are not integrated over on the RHS. These
dependencies are encoded in a SP-like dependence of the $Q$ and $q$'s
on the LHS. In addition, the RHS also depends on all the fields
contained in the delta functions $\delta_{a \to i}$ and $\delta_{a \to
  j}$ which are also present on the LHS. To give a pictorial
representation then, referring to Figure \ref{fig:2}, left panel, the
delta functions on the RHS depends on the fields represented as red
and white arrows. These dependencies are encoded on the LHS by the $Q$
and the $q$'s and the two deltas. The RHS also depends on the fields represented as black
arrows entering the purple region in the same figure. These
dependencies are encoded on the LHS by the functions $\mathcal{L}(\vec
U, \vec u)$ in the two delta functions $\delta_{a \to i}(u_{a \to
  i}-\mathcal{L}_{a \to i}(\vec U, \vec u))$ and $\delta_{a \to
  j}(u_{a \to j}-\mathcal{L}_{a \to j}(\vec U, \vec u))$. 
  
  Summarizing,
equation \eqref{eq:Qrod2} and \eqref{eq:Qsp} remain consistent
after the SP-like ansatz is taken and therefore represent a class of
generalized GBP equations for the distributions of the field messages, \emph{i.e.}
$Q(U,u,u)$ and $q(u)$, similarly as the Survey Propagation
\cite{braunstein2005survey,MezardParisi2001,MezardParisiZecchina2002,mezard2002random}
equations are Belief Propagation equations for the distribution of the
field-messages, \emph{i.e.} $q(u)$. Although these equations are consistent, they would
naturally be cumbersome to use in practice. Indeed the right-hand side is
a convolution of many terms and, more importantly, the left-hand side is a product of several distributions and delta functions. 
Despite the delta functions appearing on the LHS could be integrated on both side of the equations, the product of distributions on the LHS does not allow to use a population dynamic algorithm to iterate the equations. Therefore, the paramagnetic regime, where the equations very much simplify and the product of distribution on the LHS reduces to a single distribution $Q(u)$, seems to be the only regime where the equations can be iterated (see \cite{rizzo2010replica} for a pioneering contribution about simulations in the paramagnetic phase of the average case). 


\subsection{Comparison with the replica cluster variational method}
\label{sec:compare_rcv}

In an earlier effort Rizzo and co-authors \cite{rizzo2010replica} devised
a replica cluster variational method (CVM) for studying the EA spin glass model.
Here we discuss the relationship between our work and this
replica cluster variational approach.\\
Reference \cite{rizzo2010replica} starts applying the CVM formulation to the variational free energy 
of a $n$-times replicated system. Through a standard variational approach replica-GBP equations
are derived. The main idea of the authors is then to use a replica symmetric or a more general Parisi's 
hierarchical ansatz in order to send the number of replicas to zero and so obtain the corresponding 
RS or (RSB) GBP equations. The core of \cite{rizzo2010replica} is dedicated to the analytical and numerical
study of the paramagnetic and spin glass replica-symmetric phase of 2D EA model, by considering a plaquette approximation
of the averaged free energy. In the Appendices, the authors also show the GBP equations for a generic $k$-RSB
ansatz, both at the averaged and single instance case.  

Our method, on the other hand, starts from the assumption that each
GBP fixed point corresponds to a macroscopic state at the 1RSB level. 
We then build a factor graph model at the 1RSB level in which variable nodes
are functions (GBP messages at the RS level) and factor nodes are
functional of these GBP messages. 
The GBP equations on this new factor graph model is our 1RSB GBP scheme.
We then focus on the single instance case and therefore do not perform
averages over the quenched disorder. Generally speaking,
both the replica CVM and our approach can be in principle be used to derive RSB GBP schemes for single instances
as well as average cases. 

It is interesting to observe that these two approaches provide the same kind of equations for non-redundant region
graphs or, even more simply, tree-like topologies. When the graph considered shows redundancy, the sets of equations
provided by the two approaches are different and result to be the same only in the paramagnetic phase. We therefore 
expect, in this regime, a physical behaviour different to the one studied and encountered in \cite{rizzo2010replica}. The dissimilarities
between the two approaches mainly come from a different derivation. We list them shortly in the following.


In our approach we start assuming that parent-to-child second-level messages depend on all the $u$-field contained in the child region. This means that, plaquette-to-rod $Q$ messages depend on 14 field-messages (see dependencies in eq. \eqref{eq:Q_plaq}) whereas rod-to-vertex $q$ messages depend on 4 field-messages (see eq. \eqref{eq:Q_rod}). After the SP ansatz though, our parametrization of the messages turns out to be the same as Rizzo \emph{et al.} (see Sec. \ref{s:no-GSP}).

Second, in \cite{rizzo2010replica} the authors point out that the replica CVM $Q$-messages, 
at the RS level and in the average case, are positive definite only in the paramagnetic phase - where their equations coincide with ours - and therefore can be interpreted as populations only in this regime. As soon as 
the temperature is lowered below the spin glass transition, they observe that these messages are not
necessarily positive definite, therefore they should not be interpreted as probability functionals. In the CVM
treatment, then, only the beliefs of second-level regions have a probabilistic meaning. 
This non-positive definite property indicates that the messages at the RS levels are correlated and
redundant. In our formulation, before the SP ansatz is taken as discussed Sec. \ref{s:no-GSP}, the 1RSB  parent-to-child $Q$-messages depend on all the $u$-fields in the child region, as stated above and in eqs. \eqref{eq:Q_plaq} and  \eqref{eq:Q_rod}. This parametrization accounts for correlations among $u$-messages and we therefore believe that the $Q$'s could be considered as probability functionals. The SP ansatz taken in Section \ref{s:no-GSP} corresponds to project these functionals on a subspace in order to reduce their dimensionality. We conjecture that the non-positive definite property of the $Q$'s would only appear for the average case. In this case all the parent-to-child $Q$'s are, indeed, not region dependent and there exist only one type of plaquette-to-rod and rod-to-vertex second-level message. These latter appear in a convoluted form in the plaquette-to-rod equation and therefore the achievement of a fixed point requires to relax the positive-definite property of the $Q$'s. Such extra assumption should not be needed in the single instance case. 
 
On a technical and, more important, physical point of view, we also mention that every $k$-RSB formulation of the GBP 
equations should deal with the gauge fixing of these equations at the first-level. This issue was firstly raised in \cite{dominguez2011characterizing}
by the same authors of \cite{rizzo2010replica} and following these considerations we addressed this issue in Section \ref{Sec:GF}. The same
could be done in the CVM replica scenario but at the time of \cite{rizzo2010replica} this was not yet completely understood. 

\section{Discussion}
\label{s:discussion}
We have considered
a one-step replica symmetry breaking description of the Edwards-Anderson model in 2D through 
a second-level statistical model built on the extremal points of a 
Kikuchi approximation. These extremal points can be computed as fixed
points of Generalized Belief Propagation (GBP) in a parent-to-child message-passing
scheme on a region graph with redundancy.
We have discussed the fact that these GBP equations exhibit a gauge
invariance and how that can be corrected for by setting some parts of
the GBP messages to zero.
We have then shown that a second-level theory can be constructed 
where the variables are probability distributions over GBP messages,
and the fixed point equations are consistency equations between 
marginal probabilities of regions in the second-level region graph.
We have further shown that this theory has a set of solutions 
analogous to Survey Propagation (SP) where each second-level message
in the second-level model
depends only on the corresponding first-level message in the
first-level model. For this to be possible it is necessary to again use the
gauge invariance and gauge fixing of GBP.

Three comments impose themselves. The first concerns the resulting
Generalized Survey Propagation (GSP) equations given in final form
as equations (\ref{eq:Qrod2}) 
and (\ref{eq:Qsp}). While it would in principle be possible to
simulate them numerically, in practice that would be quite
a cumbersome task due to their high dimensionality
and the product of three messages on the left-hand side of  (\ref{eq:Qsp}).
For high enough temperatures,
within the paramagnetic phase, the above equations would simplify and numerical 
treatments become easier although still  involving. A numerical investigation of this regime, as reported for instance 
in \cite{rizzo2010replica} for the replica symmetric case, goes beyond the purpose of this paper. 
Our contribution is
therefore mainly 
of a conceptual nature, pointing to how a generalization of Survey Propagation
(SP) to finite-dimensional systems can be carried out, and how complex
such an approach appears to have to be.

A second comment concerns the determinant (Jacobian) discussed around
equation (\ref{eq:Jacobian}). To our knowledge such determinants have
never been discussed before in the literature concerning GSP. It is
clear that they must always appear as long as the second-level region graph has
loops, and it is also clear that the determinant would be a global characteristic of such a
model. On a second-level Bethe lattice, which underlie SP and which has
no loops, such determinants is equal to one.
We believe that the analysis of such determinants is an
interesting though non-trivial task in itself. 

Finally, given the prediction that a spin glass is
only present at zero temperature in the Edwards-Anderson model in 2D
then we should take the zero temperature limit.
We note that in the application of 1RSB to random satisfiability problems the
temperature is also zero,
corresponding to hard constraints, in the first
version of SP considered together with the limit of $y$ equal to zero
\cite{MezardParisiZecchina2002,mezard2002random}.
All solution clusters or zero-energy states are then weighted equally
with no entropic terms depending on the size of each cluster.
In a later development it was shown that for the same problems 
and when $y$ is larger than zero entropic terms appear that separate
clusters of different sizes \cite{mezard2009information}.
It may be that the determinant referred to above complicates the
picture in the case at hand, particularly for small $y$, and we
therefore leave this issue to future work. \\

\textbf{Acknowledgments.}
\label{s:acknowledge}
The authors warmly thank Alejandro
Lage-Castellanos and Federico Ricci-Tersenghi for valuable discussions
which helped us considerably improve upon a previous draft of the paper. Anonymous 
Referees are also acknowledged for useful comments during the reviewing process of the manuscript. 
This research is supported by 
FP7/2007-2013/grant agreement no 290038 (GDF),
by the Swedish Science Council through grant 621-2012-2982 (EA),
by the Academy of Finland through its Center of Excellence COIN (EA),
and by the Natural Science Foundation of China through
grant 11225526 (HJZ). GDF and EA thank the hospitality of KITPC and
HJZ thanks the hospitality of KTH.

\appendix


\section{A derivation of GBP}
\label{a:Kikuchi-derivation}
In this appendix we derive the GBP update equations (\ref{eq:gbp}) from a constrained variation of the Kikuchi free energy
functional (\ref{eq:F-Kikuchi}). The material is standard and included only for completeness, see~\cite{YedidiaFreemanWeiss2003}.
The task is to minimize the Kikuchi free energy functional $F_K$ in (\ref{eq:F-Kikuchi}) \textit{i.e.} 
\begin{equation}
    F_K = \sum_{\alpha}c_{\alpha} \sum_{\ul{\s}_{\alpha}} P_{\alpha}(\ul{\s}_{\alpha})   
\biggl[E_{\alpha}(\ul{\s}_{\alpha}) + \beta^{-1}\log P_{\alpha}(\ul{\s}_{\alpha}) \biggr]
\nonumber 
\end{equation}
where we have introduced the energy of a region as
\begin{equation}
    E_{\alpha}(\ul{\s}_{\alpha}) = \sum_{i\in \alpha} h_i\s_i+ \sum_{a\in \alpha} J^{(a)}_{ij}\s_i\s_j. 
\nonumber 
\end{equation}
The variation is of the marginal probabilities $P_{\alpha}(\ul{\s}_{\alpha})$   
under the constraints that they have to be consistent with each other.
This means, in the situation under consideration, that if a rod region $a$ is a child of plaquette region $\alpha$
then $\sum_{\ul{\s}_{\alpha}\setminus \ul{\s}_{a}} P_{\alpha}(\ul{\s}_{\alpha})=  P_{a}(\ul{\s}_{a})$ where 
$\ul{\s}_{\alpha}\setminus \ul{\s}_{a}$ is the exclusion set consisting of all spins in $\alpha$ that are not in
$a$. Similarly, if a vertex region $i$ is a child of rod region $a$ then 
$\sum_{\ul{\s}_{a}\setminus \s_i} P_{a}(\ul{\s}_{a})=  P_{i}(\s_{i})$. Introducing Lagrange multipliers 
$\lambda_{\alpha | a}(\s_l,\s_k)$ and $\lambda_{a | i}(\s_i)$ to represent all these constraints,
where the spin indexing follows Fig.~\ref{fig:1}, as well Lagrange multipliers $q_{\alpha}$, $q_a$ and $q_i$
for the normalizations of $P_{\alpha}$, $P_a$ and $P_i$, we arrive at
\begin{equation}
\begin{array}{lcl}
    P_i &\propto& e^{-\beta E_i} e^{\frac{1}{c_i}\beta \sum_{a:a\to i} \lambda_{a | i}} \\
    P_a &\propto& e^{-\beta E_a} e^{-\frac{1}{c_a}\beta \sum_{i:a\to i} \lambda_{a | i}} e^{\frac{1}{c_a}\beta \sum_{\alpha:\alpha\to a} \lambda_{\alpha | a}} \\
    P_{\alpha} &\propto& e^{-\beta E_{\alpha}} e^{-\frac{1}{c_{\alpha}}\beta \sum_{a:\alpha\to a} \lambda_{\alpha | a}} 
\end{array}
\label{eq:Bs-Lagrange-way}
\end{equation}
Referring to Fig.~\ref{fig:1} for the labeling of the regions, where \textit{e.g.}
vertex $i$ has four rod parents, of which one is $a$ and the one opposite is called $d_8$,
and a rod has two plaquette parents, of which one is $\alpha$ and the other one is $\beta$,
we can introduce auxiliary normalized quantities
\begin{eqnarray}
    m_{a\to i} &\propto& e^{\beta \lambda_{d_8 | i}}\\
    m_{\beta \to a} &\propto& e^{-\beta \lambda_{\alpha | a}- \lambda_{d_6 | j} - \lambda_{d_7 | i}}  
\end{eqnarray} 
In terms of the $m_{a\to i}$ and the $m_{\alpha \to a}$ the marginal probabilities of the regions then
take the GBP output equation form \textit{i.e}
\begin{equation}
\begin{array}{lcl}
    P_i &\propto& e^{-\beta E_i} \prod_{a:a\to i}m_{a\to i}\\
    P_a &\propto& e^{-\beta E_a} m_{\alpha\to a}m_{\beta\to a}\prod_{b:b\to i\setminus a}m_{b\to i}\prod_{c:c\to j\setminus a}m_{c\to j}\\
    P_{\alpha} &\propto& e^{-\beta E_{\alpha}} m_{\beta\to a}m_{\rho\to b}m_{\eta\to d}m_{\lambda\to c} \prod_{k=1}^8 m_{d_k\to \cdot}\\  
\end{array}
\label{eq:Bs-GBPway}
\end{equation}
where the dummy argument in $m_{d_k\to \cdot}$ indicates the vertex in the intersection of $d_k$ and $\alpha$.
Varying the messages in (\ref{eq:Bs-GBPway}) gives the GBP update equations as described in the main text.
Alternatively one can check that the marginalization constraints expressed in the Lagrange multipliers themselves,
which are
\begin{equation}
\begin{array}{lcl}
    e^{-\beta \sum_{b:b\to i\setminus a}\lambda_{b | i}} &\propto& \sum_{\s_j}e^{-\beta(E_a-E_i)+\beta\lambda_{a | j} -\beta\left(\lambda_{\alpha | a} +\lambda_{\beta | a}\right)}   \\
    e^{\beta\left(\lambda_{a | i}+\lambda_{a | j}\right)-\beta\lambda_{\beta | a}}\ &\propto& \sum_{\s_l,\s_k}e^{-\beta(E_{\alpha}-E_a)-\beta\sum_{c:\alpha\to c\setminus a}\lambda_{\alpha | c}}  
\end{array}
\label{eq:GBPupdate-Lagrange-way}
\end{equation}
are equivalent to the GBP update equations for the messages.
We observe that nothing in the above derivation really depends on how many variables (spins) there are in the various regions and what
their interactions are; the derivation therefore also holds for the second-level GBP.


\section{GBP equations for the 2D Edwards-Anderson model: from messages to fields.}
\label{a:detailed-GBP}
In this appendix, by parametrizing the plaquette-to-rod and rod-to-vertex messages in terms of cavity fields, we want to rewrite the GBP equations \eqref{eq:GBPplaq} and \eqref{eq:GBProds-main}  in terms of fields. As already shown in the main text, a message $m_{a \to i}(\s_i)$, which is a probability distribution on the spin $\s_i$ up to a normalization, can be written as $e^{\beta u_{a \to i}\,\s_i}$ where  $u_{a \to i}$ is the
cavity field on spin $i$ from interaction $a$. A plaquette-to-rod message $m_{\alpha \to a}(\s_i,\s_j)$ is a probability distribution on two spins and can be written, up to a normalization, as $e^{\beta\left(u_{\alpha \to a}^{(i)}\s_i+u_{\alpha \to a}^{(j)}\s_j+U_{\alpha \to a}\s_i\s_j\right)}$ where the new quantity $U_{\alpha \to a}$ parametrizes two body interactions. 
The GBP equations \eqref{eq:GBPplaq} and \eqref{eq:GBProds-main} are then relations between parameters
of the types  $u_{a \to i}$, $u_{\alpha \to a}^{(i)}$ and $U_{\alpha \to a}$. In addition we remind that the potential functions $\psi$ present in the GBP equations  are given accordingly to the definition in the EA model as in \eqref{eq:potential}, namely $\psi_i (\s_i) = e^{\beta h_i \s_i}$ and  $\psi_a(\ul\s_{\partial a}) = e^{\beta J_{ij}^{(a)}\s_i \s_j}$.\\
\textbf{Rod-to-vertex equation:}
To write this equation explicitly we note that the RHS of \eqref{eq:GBProds-main} can be parametrized as a function of the spin $\s_i$ and $\s_j$:
\be
m_{a \to i}(\s_i)=  \frac{1}{Z_{a \to i}} \exp{[\beta u_{a \to i}\, \s_i]} = \frac{1}{Z_{a \to i}}\sum_{\s_j} e^{\beta\left(x_{i}^{}\s_i+x_{j}^{}\s_j+X_{ij}^{}\s_i\s_j\right)}\, ,
\ee
where
\begin{align}
x_i &=  u_{\alpha \to a}^{(i)} + u_{\beta \to a}^{(i)}\, ,\\
x_j &= h_j +u_{\alpha \to a}^{(j)}+ u_{\beta \to a}^{(j)} + u_{c\to j} + u_{d_5 \to j} + u_{d_6 \to j}\, ,\\
X_{ij} &=  J_{ij}^{(a)} +U_{\alpha \to a } + U_{\beta \to a}.
\end{align}
Computing the rate $m_{a \to i}(1)/m_{a \to i}(-1)$ with the parametrization given above and taking the logarithm on both sides of the resulting equation, we obtain the expression for the rod-to-vertex field \cite{dominguez2011characterizing}:
\begin{equation} \label{eq:RtoV_field} 
u_{a \to i}= x_i + \frac{1}{2 \beta} \ln \Big[ \,\frac{\cosh[\beta(x_j + X_{ij})]}{\cosh[\beta(x_j - X_{ij})]} \Big]
\end{equation}
\textbf{Plaquette-to-rod equation:} In the following  we will not keep track of the normalization constants of the various messages. With the parametrization of messages in terms of fields given above, equation \eqref{eq:GBPplaq} can be written as:
\be\label{eq:fields_plaq}
\exp\Big[\beta\Big( (u_{\alpha \to a}^{(i)}+ u_{b \to i})\s_i + (u_{\alpha \to a}^{(j)}+ u_{c \to j})\s_j +U_{\alpha \to a}\, \s_i \s_j\Big) \Big] = e^{\,\,u_{\rho \to b}^{(i)}\s_i +u_{\lambda \to c}^{(j)}\s_j} K(\s_i,\s_j)
\ee
where the function $K(\s_i,\s_j)$, according to the RHS of \eqref{eq:GBPplaq}, reads as:
\begin{align} \nm
K(\s_i,\s_j) = \sum_{\s_k,\s_l} \exp\Big[&\Big( (h_l +u_{\rho \to b}^{(l)}+u_{d_1 \to l} + u_{d_2 \to l} + u_{\eta \to d}^{(l)})\s_l  
\\  &(h_k +u_{\eta \to d}^{(k)}+u_{d_3 \to k} + u_{d_4 \to k} + u_{\lambda \to c}^{(k)})\s_k   +\\ \nm
&(J_{li}^{(b)}+U_{\rho \to b})\s_l \s_i +(J_{lk}^{(d)}+U_{\eta \to d})\s_l \s_k +(J_{kj}^{(c)}+U_{\lambda \to c})\s_k \s_j  \Big)\Big].
\end{align}
After some simple algebra the plaquette-to-rod fields on the LHS of \eqref{eq:fields_plaq} are then given by
\begin{align} \label{eq:u_i}
u_{\alpha \to a}^{(i)}= & u_{\rho \to b}^{(i)}- u_{b \to i} +\frac{1}{4 \beta} \log \Big[\frac{K(1,1)K(1,-1)}{K(-1,1)K(-1,-1)}\Big]\, ,\\
\label{eq:u_j}
u_{\alpha \to a}^{(j)}=&u_{\lambda \to c}^{(j)}-  u_{c \to j}+ \frac{1}{4 \beta} \log\Big[\frac{K(1,1)K(-1,1)}{K(1,-1)K(-1,-1)}\Big] \, ,\\
U_{\alpha \to a}=&\frac{1}{4 \beta} \log\Big[\frac{K(1,1)K(-1,-1)}{K(1,-1)K(-1,1)}\Big].
\label{}
\end{align}
where we used a similar notation as in \cite{dominguez2011characterizing} to facilitate understanding. 

\section{Useful relations relative to the the partition functions}
\label{ref:useful-relations}

In this appendix we want to show some relations among the different partition functions encountered in the main text which are useful for the presentation of the theory in Sec \ref{s:GBP2}.
We wish to show that the partition functions of the plaquette, rod and vertex regions are related among each other through the partition functions which appear as normalizing constant in the GBP equations \eqref{eq:GBPplaq} and \eqref{eq:GBProds-main}. Let us write explicitly the partition functions for the region considered through the paper:
\begin{align} \nm \label{eq:Zalpha_espl}
Z_{\alpha}&=\sum_{\s_i,\s_j,\s_k,\s_l}\psi_i(\s_i)\, \psi_j(\s_j)\,\psi_k(\s_k)\, \psi_l(\s_l) \psi_a(\s_i, \s_j)\,\psi_b(\s_l, \s_i)\,\psi_c(\s_j, \s_k)\,\psi_d(\s_k, \s_l)\\ \nm
&\times m_{\beta \to a}(\s_i, \s_j) m_{\rho \to b}(\s_l, \s_i) m_{\eta \to d}(\s_k, \s_l)m_{\lambda \to c}(\s_j, \s_k)\\ 
&\times m_{d_1\to l}(\s_l) m_{d_2\to l}(\s_l)m_{d_3\to k}(\s_k)m_{d_4\to k}(\s_k)m_{d_5\to j}(\s_j)m_{d_6\to j}(\s_j)m_{d_7\to i}(\s_i)m_{d_8\to i}(\s_i)\\ \nm
Z_{a}&= \sum_{\s_i, \s_j} \psi_i(\s_i)\, \psi_j(\s_j)\, \psi_a(\s_i, \s_j) m_{\alpha \to a}(\s_i, \s_j)m_{\beta \to a}(\s_i, \s_j)\\ \label{eq:Za_espl}
&\times m_{b\to i}(\s_i) m_{d_8\to i}(\s_i) m_{d_7\to i}(\s_i)m_{c\to j}(\s_j)m_{d_5\to j}(\s_j)m_{d_6\to j}(\s_j)\\ 
\label{eq:Zi_espl}
Z_i&= \sum_{\s_i} \psi_i(\s_i)\, m_{a\to i}(\s_i) m_{b\to i}(\s_i)m_{d_8\to i}(\s_i) m_{d_7\to i}(\s_i)
\end{align}
Let us first consider the relation between the rod and vertex partition function. The rod partition function \eqref{eq:Za_espl} can be rewritten as 
\begin{align}\nm
Z_a&=  Z_{a \to i}\,\int   \sum_{\s_i} \psi_i \, m_{a\to i} \, m_{b\to i}\,m_{d_8\to i}\, m_{d_7\to i} \, \delta(m_{a\to i} - \mathcal{L}_{a\to i})\,d m_{a \to i} \\ \nm
&= Z_{a \to i}[m_{\alpha \to a}, m_{\beta \to a} m_{c \to j},  m_{d_5 \to j},  m_{d_6 \to j}]\\
&\times\int Z_i(m_{a\to i},m_{b\to i}\,m_{d_8\to i}\, m_{d_7\to i})  \, \delta(m_{a\to i} - \mathcal{L}_{a\to i})\,d m_{a \to i} 
\label{eq:Zai}
\end{align}
where in the first equality we used the RHS of the GBP equation \eqref{eq:GBProds-main} to recognize $m_{a \to i}$ in \eqref{eq:Za_espl} and we introduced the delta function to enforce such a GBP equation. The function $Z_{a \to i}$ is the constant appearing in equation \eqref{eq:GBProds-main}, its and the $Z_i$'s dependencies on the messages are written explicitly on the RHS for clarity. Using the expression \eqref{eq:Zalpha_espl} and \eqref{eq:Za_espl}, a similar calculation shows that the plaquette partition function $Z_{\alpha} $ can be written as
\begin{align}
Z_\alpha &= Z_{\alpha \to a}[m_{\rho \to b},m_{\eta \to d},m_{\lambda \to c},m_{d_1 \to j}, m_{d_2 \to j}, m_{d_3 \to k}, m_{d_4 \to k}]\int Z_a \, \delta(m_{\alpha \to a}- \mathcal{F}_{\alpha \to a})\, d m_{\alpha \to a}
\label{eq:Zalphaa}
\end{align}
where we used the the RHS of the GBP equation \eqref{eq:GBPplaq} to recognize and substitute terms equivalent to $m_{\alpha \to a}$ in \eqref{eq:Zalpha_espl} and the delta function to enforce such a GBP equation. The function $Z_{\alpha \to a}$ above corresponds to the constant factor of equation \eqref{eq:GBPplaq} with its own dependencies written explicitly for clarity.

\section{Computation of determinants}
\label{s:determinants}

As already pointed out in the main text, a proper formulation of the second-level GBP equations should take into account determinant factors which act as entropic corrections in the re-weighting terms included in the partition functions at the second-level. 

We encountered two classes of these corrections, the first one is the determinant appearing in equation \eqref{eq:Jacobian}. This term corresponds to a large $L \times L$ matrix  where $L$ is the number of GBP messages (plaquette-to-link and link-to-vertex) appearing in the whole square lattice considered. Its contribution can be written as $ | \textit{\rm{det}}(\mathds{1}-\mathds{J})| $ where $\mathds{J}$ is the $L \times L$ matrix of all the derivatives of the GBP equations respect to all the plaquette-to-link and link-to-vertex field messages computed at the fixed point. Its explicit computation is a cumbersome task therefore, at a first level of modelling, we neglected this contribution. 
\begin{figure}[!t]
  \begin{center}
    \includegraphics[width=1.6cm]{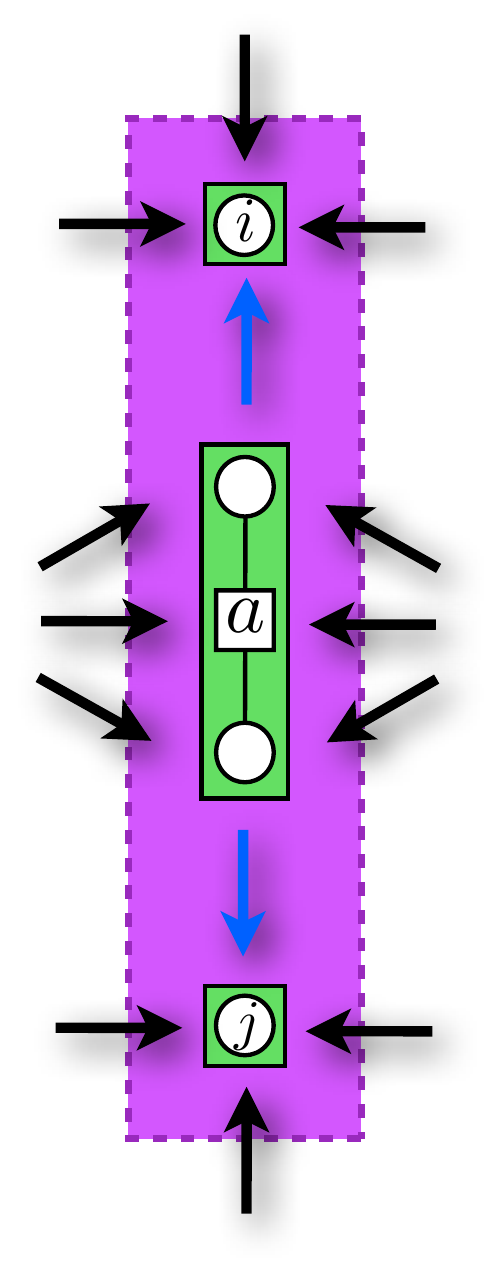}\hspace{1.5cm}
    \includegraphics[width=4cm]{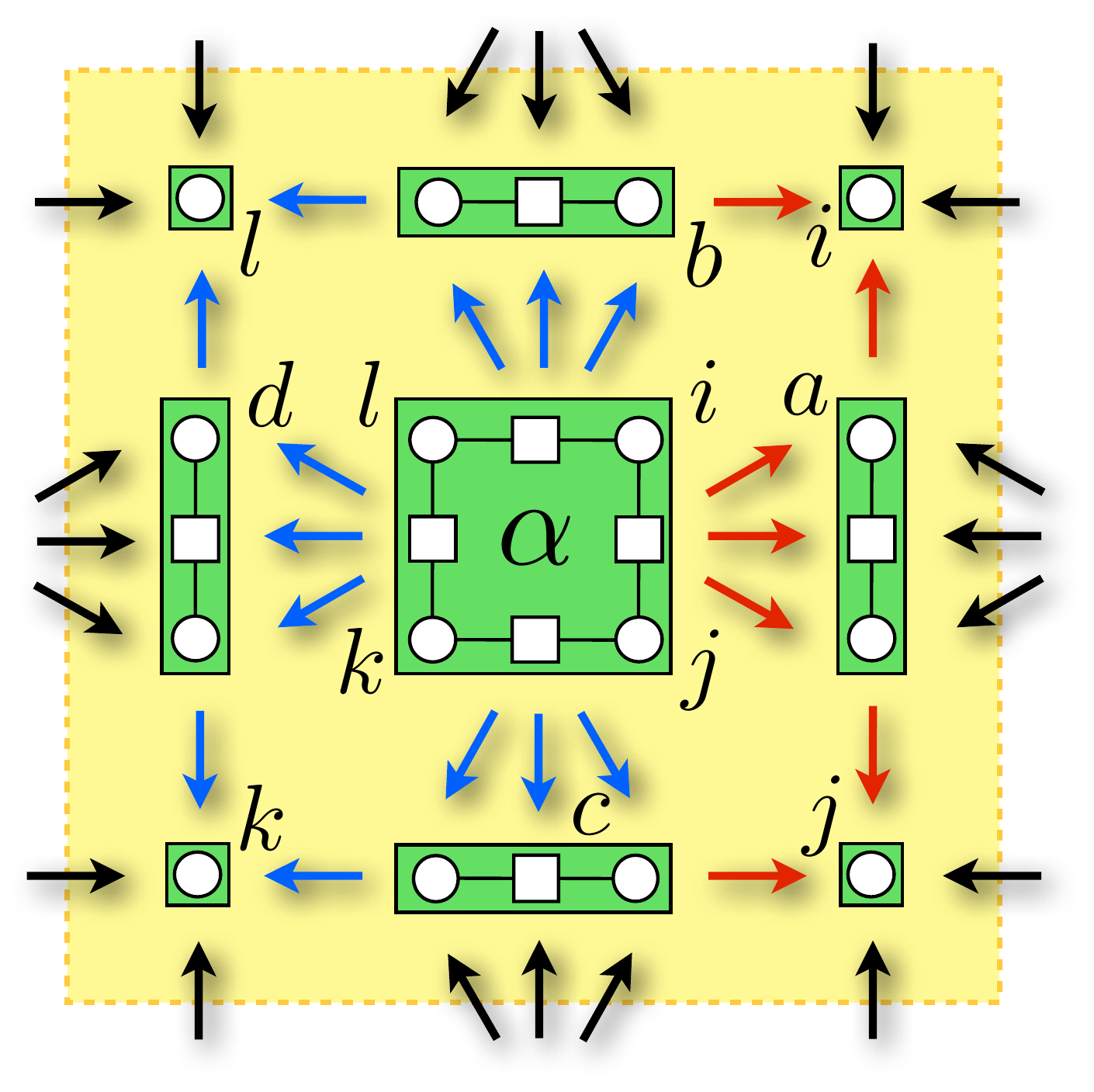}\hspace{1.5cm}
    \includegraphics[width=4cm]{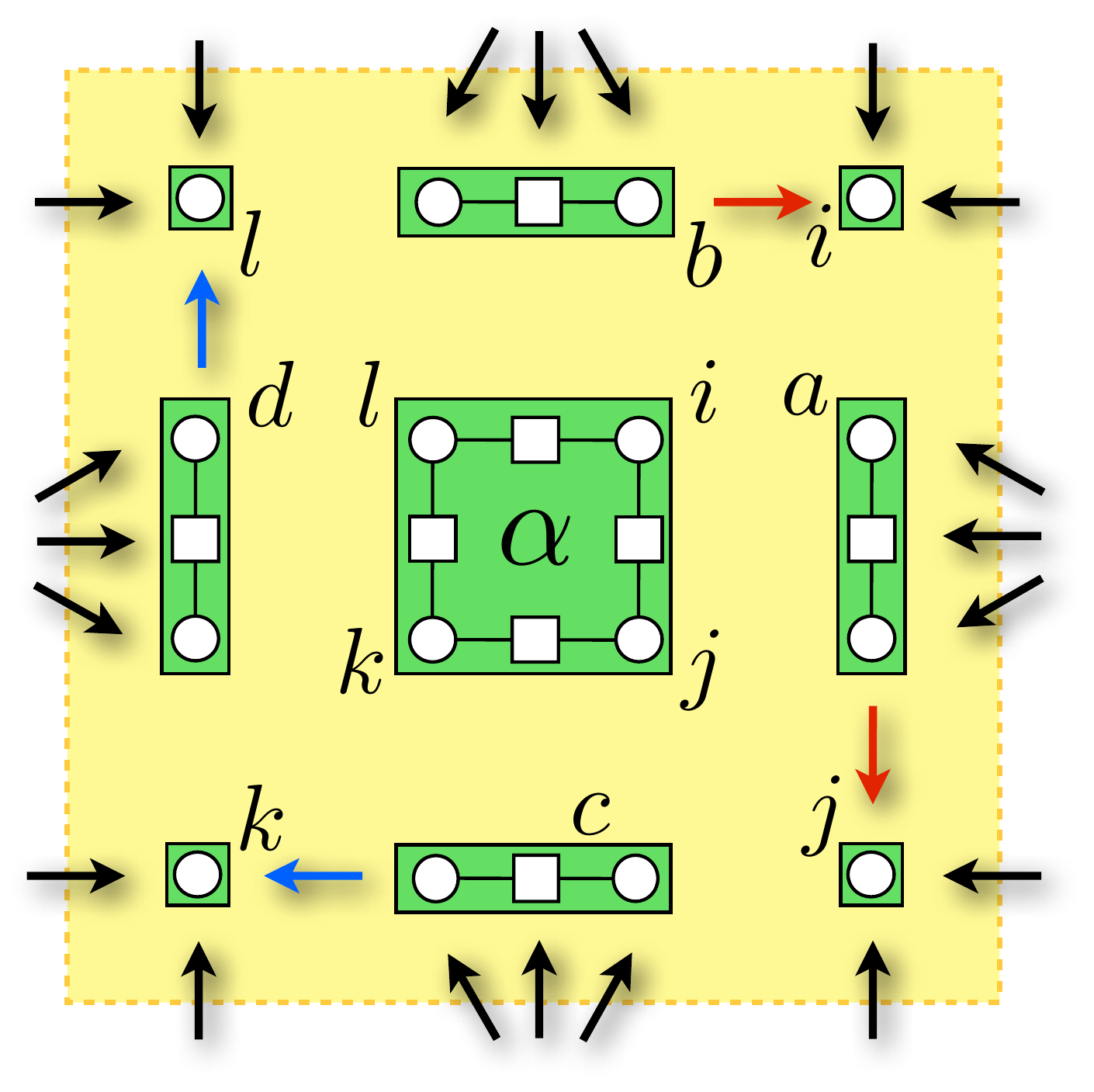}
    \caption{\footnotesize Left and central panel: second level rod and plaquette region respectively with external field-messages depicted as black arrows and internal depicted either as blue or red ones. Right panel: plaquette region after some of the internal field-messages have been integrated out. All the red arrows in the central and right panel represent fields with gauge fixed (see Section \ref{Sec:GF} and \ref{sec:ConsEqPR}).}
    \label{fig:det}
  \end{center}
\end{figure}

We can instead compute explicitly the second class of determinant corrections arising in the main text, which are those appearing in the consistency equations \eqref{eq:cons_rod_vert} and \eqref{eq:GBP2-plaquette-to-rod}, used to derive the GBP equations at the second-level (see lines below these equations). In the main text we claim that these two determinants are equal to one. Hereby, rather than computing explicitly the matrix of all the derivatives in the two cases and calculate it at the fixed point, we give a simple argument which proves our claim. 

Let us recall that determinants appear in the equations because of the integration of the delta functions, we here show that, these integrations can be carried out completely without entering in loops and, therefore, the determinant contribution is equal to one. Hereafter we refer to figure \ref{fig:det} for a pictorial representation. For the case of the equation \eqref{eq:cons_rod_vert}, the term $| \textit{\rm{det}}(\mathds{1}-\Delta \mathcal{L})|$ rises because of the presence of the two delta functions $\delta_{a\to i}$ and $\delta_{a\to j}$. The fields enforced by these deltas are represented in figure \ref{fig:det}, left panel, by blue arrows which (according to their GBP equations) only depends on the external fields of the link region (depicted as black arrows). Therefore the integration over $u_{a \to i}$ and $u_{a \to j}$ can be carried out and the result will be just the change of the value of these two fields with external ones and hence the determinant contribution turns out to be one. 

More involved is the same proof for the term $| \textit{\rm{det}}( \mathds{1}-\Delta \mathcal{F})|$ appearing in equation \eqref{eq:GBP2-plaquette-to-rod}. This determinant emerges because of the integration over all the delta functions enforcing GBP conditions for the fields in $I(\alpha)$, depicted with (non-black) arrows in figure \ref{fig:det}, central panel. 

We here want to stress that, to finally prove that also this term in equal to one, it is crucial to fix the gauge (see Section \ref{Sec:GF} and \ref{sec:ConsEqPR}), \textit{i.e.} to change some deltas from $\delta(u_{x \to y} - F_{x \to y}(\dots))$ to $\delta(u_{x \to y})$. This indeed breaks the loop between the GBP equations inside the plaquette region, as it will be clear in the following. Fields which have been changed accordingly to the gauge fixing are depicted as red arrows in the figure above for clearness. 

The evidence is based on a similar procedure encountered in Section \ref{sec:ConsEqPR} where we integrate some delta functions to simplify the equation therein. We start integrating out all the fields $U_{\alpha\to a},U_{\alpha\to b},U_{\alpha\to c},U_{\alpha\to d}$ which are the central arrows from the plaquette-to-link in Fig. \ref{fig:det}, central panel, noting that, according to their GBP equations, these fields depend only on other external fields (black arrows). We then integrate out all the (right oriented)-fields of the type $u_{\alpha \to a}^{(j)}$ (namely $u_{\alpha \to a}^{(j)},u_{\alpha \to b}^{(i)},u_{\alpha \to c}^{(k)},u_{\alpha \to d}^{(l)}$ ) and observe that, as mentioned in the main text,  these integrations (see equation \eqref{eq:u_j}) have the final effect of changing the dependence of the (counter-clockwise oriented) fields like $u_{a \to i}$ on only external fields (see equation \eqref{eq:RtoV_field}). We then can carry out  the integration over these latter fields, namely $u_{a\to i},u_{b\to l},u_{d\to k},u_{c\to j}$. 
Once this is done, the remaining fields $u_{\alpha \to a}^{(i)},u_{\alpha \to b}^{(l)},u_{\alpha \to c}^{(j)},u_{\alpha \to d}^{(k)}$ also only depends on the external other fields (see equation \eqref{eq:u_i} considering that messages of the type $u_{b\to i}$ have been integrated out and carry only a dependence on the external fields). 

We are then left with a picture illustrated in Figure \ref{fig:det}, right panel, where there remain only four internal fields to be integrated out. As we can see, in principle, these fields are in loop chain dependence according to their GBP equations. Indeed, using $\rightarrow$ as a symbol of dependence, we have that $u_{b\to i} \rightarrow u_{d\to l} \rightarrow u_{c\to k} \rightarrow u_{a\to j} \rightarrow u_{b\to i}$ and so on. It is therefore here that, the gauge fixing condition becomes vital to achieve the result. Indeed, after the gauge is fixed, the fields depicted as red arrows in figure \ref{fig:det}, right panel, do not share any dependence on other fields and therefore can be integrated out. The field $u_{c \to k}$ hence only depends on external fields and can also be integrated out, leaving $u_{d \to l}$ depending only on external messages and therefore its integration can also be carried out with no consequences. 

This proves that the term $| \textit{\rm{det}}( \mathds{1}-\Delta \mathcal{F})|$ is also equal to one as $| \textit{\rm{det}}(\mathds{1}-\Delta \mathcal{L})|$ and therefore there are no determinant corrections to the GBP equations on a second-level. A similar argument to the one used here can be applied to show that also the determinant which appear in the SP equation in \cite{mezard2009information,zamponi2010mean} is actually equal to one. Indeed on tree topologies loops are absent and the integration over the delta functions can be carried out with no side-effects. 
%
%
%
%
%
%
\bibliographystyle{unsrt}
\bibliography{general}
\end{document}